\pdfoutput=1
 \documentclass[final,3p,times,twocolumn,sort&compress,number]{elsarticle}

\usepackage{graphicx}
\usepackage{caption}
\usepackage{subcaption}
\graphicspath{{./figs/}}

\usepackage{import}
\usepackage{epsfig}
\usepackage[bookmarks=false]{hyperref}
\usepackage[capitalise,english]{cleveref} 
\usepackage{paralist}

\usepackage{amssymb}
\usepackage{amsmath}

\usepackage{framed} 
\usepackage{multicol} 

\usepackage[noprefix]{nomencl} 
\makenomenclature
\renewcommand*\nompreamble{\begin{multicols}{2}}
\renewcommand*\nompostamble{\end{multicols}}

\usepackage{numprint}
\npthousandsep{,}\npthousandthpartsep{}\npdecimalsign{.}

\usepackage{lineno}

\renewcommand{\vec}{\boldsymbol}

\newcommand{\via}{via} 
\newcommand{\eg}{e.g.} 
\newcommand{\ie}{i.e.} 
\newcommand{\lFPZ}{$\ell_{\mathrm{FPZ}}$} 
\newcommand{\lc}{$\ell_{\mathrm{c}}$} 

\makeatletter
\newcommand*{\rom}[1]{\expandafter\@slowromancap\romannumeral #1@}
\makeatother

\makeatletter
\def\ps@pprintTitle{%
   \let\@oddhead\@empty
   \let\@evenhead\@empty
   \def\@oddfoot{\reset@font\hfil\thepage\hfil}
   \let\@evenfoot\@oddfoot
}
\makeatother

\begin{document}

\begin{frontmatter}



\title{Studying the influence of inclusion characteristics on the characteristic length involved in quasi-brittle materials using the lattice element method}


%

\author{Huu Phuoc Bui\corref{cor1}}
\ead{hphuoc.bui@gmail.com}
\cortext[cor1]{Corresponding author:\\ \indent \indent \textit{Tel.}: +33 6 47 13 09 98\\ \indent \indent \textit{Fax}: +33 4 76 82 70 43}
\author{Vincent Richefeu}
\author{Fr\'{e}d\'{e}ric Dufour}
\address{Univ. Grenoble Alpes, 3SR, F-38000 Grenoble, France}
\address{CNRS, UMR 5521 3SR, F-38000 Grenoble, France}

\begin{abstract} 
Unlike nonlocal models, there is no need to introduce an internal length in the constitutive law for lattice model at the mesoscopic scale. Actually, the internal length is not explicitly introduced but rather governed by the mesostructure characteristics themselves. The influence of the mesostructure on the width of the fracture process zone which is assumed to be correlated to the characteristic length of the homogenized quasi-brittle material is studied. The influence of the ligament size (a structural parameter) is also investigated. This analysis provides recommendations\slash warnings when extracting an internal length required for nonlocal damage models from the material mesostructure.

\end{abstract}

\begin{keyword}
Quasi-brittle materials \sep characteristic length \sep internal length \sep fracture \sep Lattice Element Method.


\end{keyword}

\end{frontmatter}


\begin{table*}[!t]
  \begin{framed}
    \printnomenclature
  \end{framed}
\end{table*}

\section{Introduction}

Fracture of quasi-brittle materials is characterized by a zone with a finite size around and ahead the crack tip, in which damage occurs and causes the softening behavior of the materials. This is the fracture process zone (FPZ)\nomenclature{FPZ}{the Fracture Process Zone}. For instance for concrete, the size (width) of the FPZ, denoted by \lFPZ{} hereafter\nomenclature{\lFPZ}{the size (width) of the fracture process zone}, is believed to be proportional to the maximum aggregate size $d_{\max}$, see, \eg,~\cite{Bazant.1983.155,PijaudierCabot.1987.1512}. Therefore, in nonlocal models (gradient or integral form~\cite{PijaudierCabot.1987.1512,Giry.2011.3431,PEERLINGS.1996.3391}), the FPZ size which only depends on the internal length \lc{} introduced, depends on (is proportional to) the maximum aggregate size. Accordingly, neither loading nor structural effect is considered to affect the resulting size of the FPZ except in the latest integral nonlocal model proposed in~\cite{Giry.2011.3431}. In the latter, the internal length parameter evolves depending on the stress state during the damage process and also depends on the intrinsic (characteristic) length that can be correlated with aggregate size of the material. However, the correlation between the characteristic length and the aggregate size has not been explicitly calibrated yet. 

The literature often reports a linear or affine relation between \lc{} and $d_{\max}$, see, \eg,~\cite{Otsuka2000,Bazant.1989.115}. But actually, varying $d_{\max}$ in experiments may lead to a number of changes in the aggregate structure characterized by other parameters such as the volume fraction of aggregate, their size distribution, their fabric or connectivity. Basic questions may be raised: what does affect the internal length of a nonlocal model? Is it only the maximum size of aggregates or some less obvious parameter(s)? Does the structure itself (size or ligament) play a role in the internal length?

To address these questions, numerical simulations of uniaxial tensile tests are carried out using the lattice model in which the geometry and mechanical properties of the material mesostructure are explicitly introduced. The output of the simulations is the FPZ size and the characteristic length of the material. The characteristic length is \textit{a priori} regarded as the internal length that would be introduced in nonlocal models. The same notation \lc{} is thus used in the following. From the lattice simulations, the relationship between the two lengths \lFPZ{} and \lc, and some relevant characteristics of the material mesostructure is found out. The study is restricted to the case of two-dimensional analysis of a brittle elastic model material with circular inclusions and is also restricted to mode-\rom{1} failure problems occurring with small deformations under quasi-static loading conditions.

It is important to stress before reading the following that the inclusions and matrix have a brittle elastic behavior together with highly simplified geometry. As a consequence, our observations and conclusions must be translated with caution to the case of real concrete.

The lattice model used in our study is briefly recalled. The model is implemented in our self-writing code using C\texttt{++} programing language. The method to assess the FPZ size and the characteristic length of the material will be next pointed out before performing numerous numerical experiments to study the influence of the material mesostructure and of the structural parameter (ligament size) on these lengths.

\section{Numerical model}

The lattice element method (LEM) is a convenient way to model the fracturing of quasi-brittle materials for the problems in which the discontinuities are dominant since it provides a discrete representation of material disorder and failure. By using the LEM, the micro-cracking, crack branching, crack tortuosity and bridging of quasi-brittle materials can easily be identified and captured. It allows the fracture process to be followed until complete failure. There exist two different types of lattice models. The first one is called classical lattice models in which the material is discretized as a network of discrete 1D-elements that can transfer forces and possibly moments~\cite{Schlangen.1992.25.153,Schlangen1992105,Schlangen1993,vanMier.1995.201}. The second type of lattice models, called particle lattice models, are classified as a discrete element method~\cite{Kikuchi1992} in which the material is discretized as an assemblage of rigid particles interconnected along their boundaries through normal and shear springs~\cite{Kawai1978}. The models in this category also include the rigid-body-spring networks~\cite{BolanderJr2000}, bonded-particle model~\cite{Potyondy1996}, random particle models~\cite{Bazant.1990.116}, beam-particle model \cite{Addetta.2002.4,Delaplace2005}, confinement-shear lattice model~\cite{Cusatis2003}. The main advantage of particle lattice models with respect to classical lattice models is that they account for the fact that crack surfaces may act on each other causing the repulsive force during the loading process. So the particle lattice models are more suitable for predicting the failure behavior in mode \rom{2} or mode \rom{1} under cyclic loadings whereas the classical ones are enough when the mode \rom{1} failure prevails.

In this work, only the mode-\rom{1} failure of the material submitted to monotonic mechanical loadings is considered. Moreover, for studying the influence of the material mesostructure on the FPZ which is related to the characteristic length of the material, a detailed description of tortuous crack patterns is important. Therefore, a lattice model, based on the classical lattice models, in which the normal and shear springs are introduced.

\begin{figure}[!htbp]
\begin{center}
\includegraphics[width=\columnwidth]{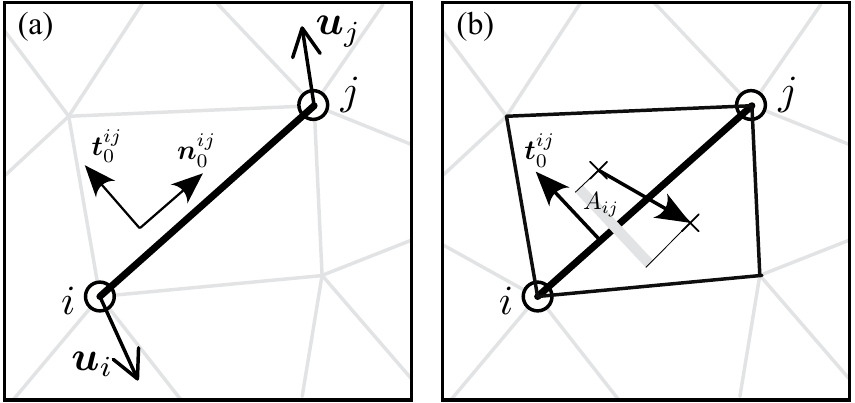}
\caption{1D-element with its local coordinate system (a) and its effective width $A^{ij}$ (b).}
\label{fig:local_schem}
\end{center}
\end{figure}

The constitutive laws of the 1D-elements are simple elastic relations in the normal and tangential directions defined by each element, see \Cref{fig:local_schem}a. Only small perturbations are considered, the positions of the lattice nodes are assumed fixed and unknown variables are the node displacements $\vec{u}$\nomenclature{$\vec{u}$}{The node displacement vectors}. The axial direction $\vec{n}_0^{ij}$ and transverse direction $\vec{t}_0^{ij}$ associated with each element $ij$ remain thus fixed. Length variations between the node $i$ and $j$ are defined by $\delta_n^{ij} = (\vec{u}_i - \vec{u}_j)\cdot \vec{n}_0^{ij}$ and $\delta_t^{ij} = (\vec{u}_i - \vec{u}_j)\cdot \vec{t}_0^{ij}$ for the normal and tangential directions, respectively. The forces are related to this length variations by $f_n^{ij} = K_n^{ij} \delta_n^{ij} $ and $f_t^{ij} = K_t^{ij} \delta_t^{ij}$, where $K_n^{ij}$ and $K_t^{ij}$ are the normal and shear stiffnesses of the element, respectively.

The approach consists in finding the set of node displacements $[\vec{u}]$ -- among which some are imposed along the boundaries -- that minimize the total elastic energy of the system:
\begin{equation}
\mathcal{U}_e([\vec{u}]) = \frac{1}{2} \sum_{ij} 
\left \{ 
K_n^{ij} (\delta_n^{ij})^2 
+ K_t^{ij}(\delta_t^{ij})^2  
\right \}
\end{equation}
\nomenclature{$\mathcal{U}_e$}{the total elastic energy of the system}
%
To proceed this minimization, the conjugate gradient method is used with the following definition of the gradient:
\begin{equation}
\frac{\partial \mathcal{U}_e}{\partial u_i^\alpha} = 
- \vec{e}_\alpha \cdot \sum_{j,\ i \in ij} 
\left \{ 
K_n^{ij}  \delta_n^{ij}  \vec{n}_0^{ij}
+ K_t^{ij} \delta_t^{ij} \vec{t}_0^{ij} 
\right \}
\end{equation}
where $\vec{e}_\alpha$ stands for the two directions of the global frame. 

The damage (in the form of diffuse or macroscopic cracks) of the whole lattice system is accounted for by removing each element that breaks according to a criterion $\psi(f_n^{ij},\ f_t^{ij}) \geq 0$. The Mohr-Coulomb surface with a cut-off of the tensile strength~\cite{BolanderJr1998569} can be adopted. However, we chose to use another model that has the advantage of being more generic while it is expressed in a single function:
\begin{equation}
\psi(f_n^{ij},\ f_t^{ij}) = 
\frac{f_n^{ij}}{A^{ij}\sigma_n^0} + \left ( \frac{f_t^{ij}}{A^{ij} \sigma_t^0} \right )^n - 1
\end{equation}
\nomenclature{$\sigma_n^0$}{the ultimate stress for pure normal loading}
\nomenclature{$\sigma_t^0$}{the ultimate stress for pure tangential loading}
where $\sigma_n^0$ and $\sigma_t^0$ are the ultimate stresses for pure normal and tangential loadings, respectively; $n$ is a positive parameter that changes the yield surface from a linear form ($n=1$) -- corresponding to the classical Mohr-Coulomb criterion -- to a non-linear form ($n>1$). In this study, $n=5$ is used.

Let us now consider a system of lattice elements where small displacements are imposed for some nodes on the boundary. A reference solution $[\vec{u}_{\rm ref}]$, corresponding to the free displacements of the other nodes, can be found by minimizing $\mathcal{U}_e$ as described above. Provided that the elements remain elastic and intact, any other elastic state is an \textit{uniform scaling} of the reference solution: $[\vec{u}] = \eta [\vec{u}_{\rm ref}]$. As a consequence, elastic forces can be scaled by the same factor and it becomes possible to find, for each element, a factor $\eta^{ij}$ so that $\psi(\eta^{ij} f_n,\ \eta^{ij} f_t) = 0$. The state corresponding to the failure of the weakest element can thus be obtained by scaling the reference solution by the factor $\eta_{\min} = \min_{ij} \{ \eta^{ij} \}$, and then recorded. The next loading state will result from another reference solution beginning from a new configuration into which the broken element is removed.   
By repeating this procedure for each element failure, one by one, the loading course is controlled by these events, rather than a time-stepping which could involve more than one element removal within a single time step. This would results in non-physical solutions that make the mechanical response dependent on the loading magnitude \cite{Delaplace2007}.

With the LEM, heterogeneities appear \textit{de facto} at the mesh level. The required disorder in the mesh, introduces a variation in  lengths $\ell^{ij}$ \nomenclature{$\ell^{ij}$}{length of the element $ij$} and effective width $A^{ij}$ of the elements \nomenclature{$A^{ij}$}{effective width of the element $ij$}. It results in an unwanted parasitic heterogeneity in the stiffness properties that can be limited by accounting for the local geometry in the element behavior:
\begin{equation}
K_n^{ij} = \frac{A^{ij}}{\ell^{ij}} \bar{K}_n^{\varphi} \quad{\rm and}\quad K_t^{ij} = \frac{A^{ij}}{\ell^{ij}} \bar{K}_t^{\varphi} 
\end{equation}
\nomenclature{$K_n^{ij}$}{the normal stiffness of the element $ij$}
\nomenclature{$K_t^{ij}$}{the tangential stiffness of the element $ij$}
\nomenclature{$\bar{K}_n^{\varphi}$}{the normal stiffness that can be set ``uniformly'' to a phase $\varphi$}
\nomenclature{$\bar{K}_t^{\varphi}$}{the tangential stiffness that can be set ``uniformly'' to a phase $\varphi$}
where $\bar{K}_n^{\varphi}$ and $\bar{K}_t^{\varphi}$ are the stiffnesses that can be set ``uniformly'' to a phase $\varphi$. The effective width $A^{ij}$ is the distance between centroids $C_{ijk}$ and $C_{ijm}$ of the triangles adjacent to the element $ij$, projected onto the local direction $\vec{t}_0^{ij}$ as proposed in \cite{Cusatis20067154}, see \Cref{fig:local_schem}b. Since the state of plane stress or plane strain is not explicitly defined in LEM-based simulations, the quantity $A^{ij}$ can also be regarded as a surface by assuming an unit length in out-of-plane direction. In this picture, $\bar{K}_n^{\varphi}$ and $\bar{K}_t^{\varphi}$ have a dimension of material stiffness. As a consequence of the weighting of imposed stiffnesses (or modulii) $\bar{K}_n^{\varphi}$ and $\bar{K}_n^{\varphi}$ in a phase, actual stiffnesses of elements differ from each other.

The targeted Young's modulus $E^{\varphi}$ and Poisson's ratio $\nu^{\varphi}$ of the phase $\varphi$ can be used to determine the element stiffnesses by the following relations:
\begin{equation}
\bar{K}_n^{\varphi} = \frac{E^{\varphi}} {1-\nu^{\varphi}} 
\quad {\rm and}\quad
\bar{K}_t^{\varphi}  = \frac{E^{\varphi}(1-3\nu^{\varphi})} {1-(\nu^{\varphi})^2 }
\end{equation}
\nomenclature{$E^{\varphi}$}{the Young's modulus of the phase $\varphi$}
\nomenclature{$\nu^{\varphi}$}{the Poisson's ratio of the phase $\varphi$}
These relations are derived from the equations given in \cite{Chang.2002.1941} for a regular and triangular lattice, by replacing a factor $\sqrt{3}$ by $1$ (found empirically from a number of single-phase simulations).

\begin{figure}[!htbp]
\centering
\includegraphics[width=0.6\columnwidth]{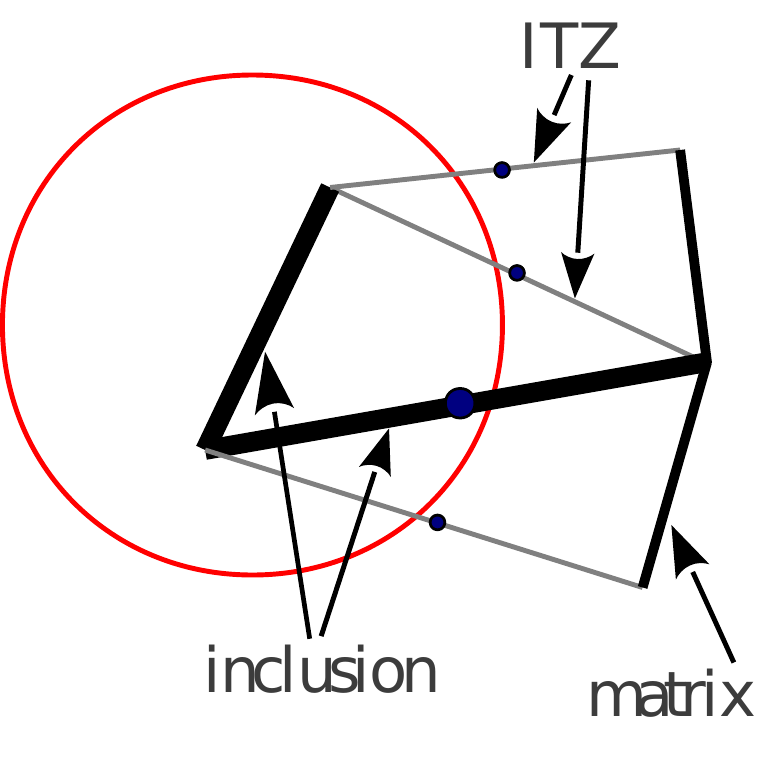}
\caption{Distinction between inclusion, matrix and ITZ phases according to the location of a lattice element in the grain structure.}
\label{fig:phaseDefinition}
\end{figure}

From there, heterogeneity intrinsic to the mesh geometry is limited as much as possible, and a structure of inclusions (grains) can be generated using the take-and-place processes~\cite{Wang_1999_533,Hafner2006}. After generating the inclusion structure, different material phases are defined and different local mechanical properties are assigned to the elements falling in each phase. At the mesoscale, three phases can be distinguished: inclusion, matrix and interfacial transition zone (ITZ) \nomenclature{ITZ}{the interfacial transition zone}, see \Cref{fig:phaseDefinition}. If both ends of an element are located in the same phase, then this element is assigned the same mechanical properties of the corresponding phase (inclusion or matrix), otherwise it is considered as interface or inclusion element depending on the location of its midpoint. If its midpoint is located within the grain, the element is classified as inclusion element, or else it will be ranked as ITZ element. The reason for this definition of ITZ element is that the resulting fraction of inclusions (the ratio between the number of inclusion elements and the total number of elements) is closer to desired fraction of inclusions in material than those developed by other authors~\cite{Schlangen.1992.25.153,Lilliu.2003.927,Sagar.2009.865}. In their models, all elements that connect two different zones of grain structures are considered as ITZ elements.

%

\section{Assessment of characteristic length}


To account for damage in continuous (and homogenized) modeling of concrete, a length parameters is required \cite{PijaudierCabot.1987.1512,PEERLINGS.1996.3391}. This length, denoted by \lc, called \textit{characteristic length} is seen as an intrinsic property of the material, however it is not so simple to determine and to connect with the heterogeneities at lower scales. The method proposed in \cite{Bazant.1989.115} is used here to assess this characteristic length for a material modeled by lattice elements. The basic idea is that the characteristic length of the material is approximated by the \textit{effective} width $h$ of the zone in which the fracture energy of the material is dissipated. This effective width is defined as the ratio of the fracture energy $G_f$ (energy per unit area of crack surface) dissipated by the cracking that localizes in a narrow band of the specimen in localized tensile test and the \textit{energy density} $W_s$ dissipated by the cracking that is nearly homogeneously distributed in the whole volume of the specimen of the same material in distributed tensile test.
Finally, the characteristic length is approximated by $h$ which can be assessed by
\begin{equation}
\ell_{\mathrm{c}} \simeq h = \frac{G_f}{W_s}
\end{equation}
\nomenclature{$\ell_{\mathrm{c}}$}{the characteristic length}
\nomenclature{$G_f$}{the fracture energy (energy per unit area of crack surface)}
\nomenclature{$W_s$}{the dissipated energy density}

\begin{figure}[!htbp]
\begin{center}
\includegraphics[width=\columnwidth]{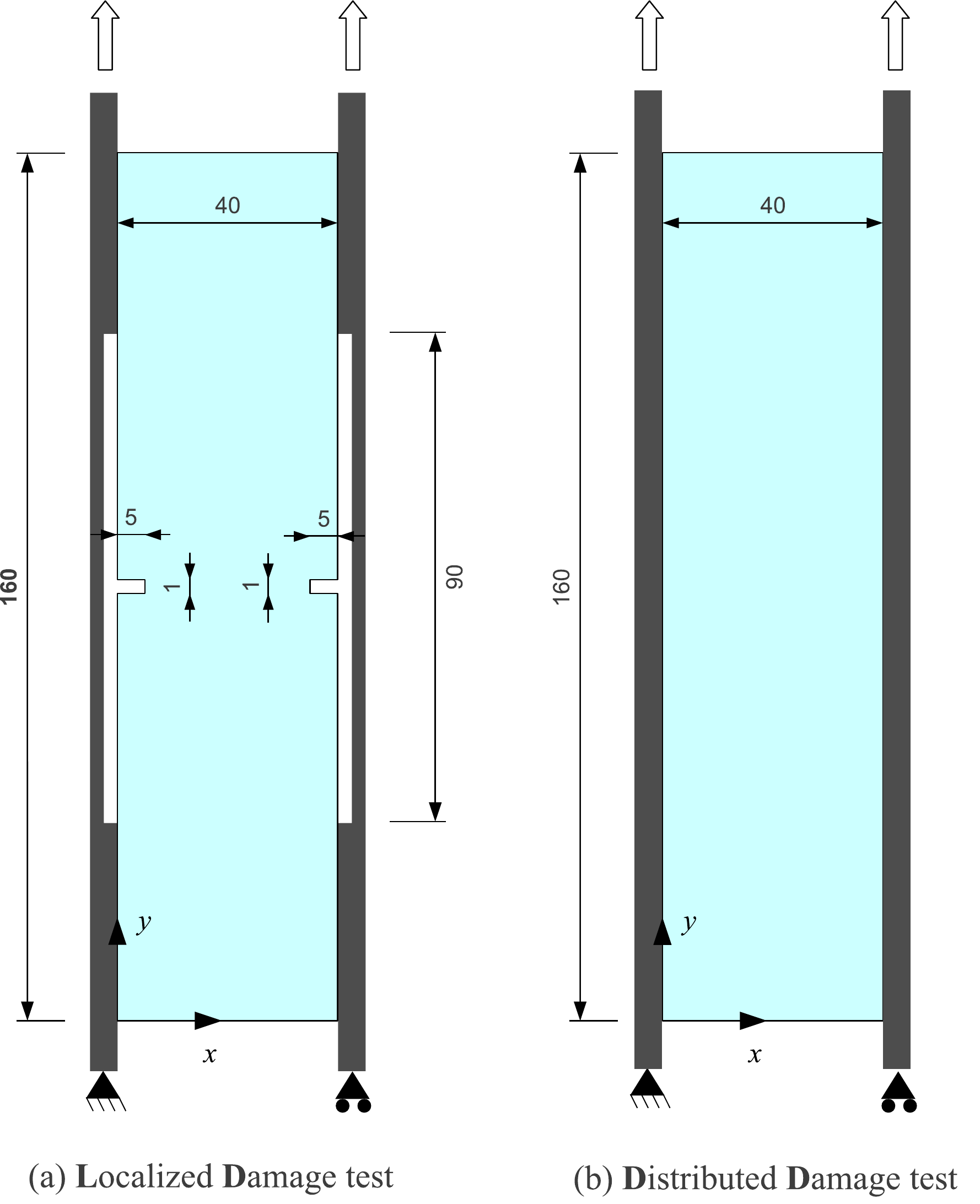}
\caption{Sketch of the specimens used to determine the characteristic length as proposed in \cite{Bazant.1989.115}. The tensile test performed on the notched specimen to obtain the localized cracking process (a) and on the unnotched specimen to obtain the distributed cracking process (b).}
\label{fig:TD_PIED_40x160}
\end{center}
\end{figure}

To evaluate \lc{} with LEM simulations, both numerical tensile tests (localized and distributed) have to be performed to determine $G_f$ and $W_s$. $G_f$ is determined from the tensile test performed on a notched specimen so that the damage can be localized whereas $W_s$ has to be determined from the tensile test carried out on an unnotched specimen with specific design of loading such that the damage is as homogeneously as possible distributed in the specimen volume. To this end, the numerical simulations of tensile tests using the lattice model can be performed in which the tensile loading is indirectly applied to the notched and unnotched specimens by elongating the steel bars ``glued'' to the specimens as proposed in~\cite{Bazant.1989.115}, see~\Cref{fig:TD_PIED_40x160}. These two tests were performed on numerical specimens of the same size, the loading is applied by means of lateral bars that are ``glued'' to the specimen and set 10 times stiffer than the material tested. The main difference between the two types of tensile tests is that the steel bars are only glued to the ends of the notched specimen within a certain length while they are entirely glued to the unnotched specimen within the whole height of the specimen. For the following, the tensile tests performed on notched specimens, where the \textbf{L}ocation of \textbf{D}amage is forced, are referred to as LD-tests. The tensile tests performed on unnotched specimens, designed to identify \textbf{D}istributed \textbf{D}amage, are mentioned as the DD-tests. These tests are known as the PIED (Pour Identifier l'Endommagement Diffus) \nomenclature{PIED}{Pour Identifier l'Endommagement Diffus} tests in the French community, as introduced in~\cite{Fokwa1992}. Note however that a diffuse damage is actually not achievable, that is why we prefer to talk about distributed rather than diffuse damage. In the lattice simulations, the steel bars with the width of $2$~mm are also discretized in 2D by the lattice elements (2D mesh) but their stiffnesses are set 10 times greater than those of the material tested and they always have an elastic behavior. The steel bars are perfectly ``glued'' to the specimens \via{} compatible nodes.

In LD-tests (\Cref{fig:TD_PIED_40x160}a), a crack is initiated and then propagates until the specimen breaks. The fracture energy $G_f$ is simply the sum of all elastic energy dissipated by the rupture of each element $ij$ divided by the total cracking surface:
\begin{equation}
\label{eq:G_f}
G_f = \frac{1}{2}  \frac{\sum_{ij} A_{ij}^2 \left ( (\sigma_n^{0})^2/K_n^{ij} + (\sigma_t^{0})^2/K_t^{ij}  \right ) } {\sum_{ij} A_{ij}}
\end{equation}

The DD-tests (\Cref{fig:TD_PIED_40x160}b) aims to avoid any onset of crack so that the straining and damage are as uniform as possible. The energy density $W_s$ is thus given by the total elastic energy dissipated within the specimen volume $V$:
\begin{equation}
\label{eq:W_s}
W_s = \frac{1}{2V} \sum_{ij} A_{ij}^2 \left ((\sigma_n^{0})^2/K_n^{ij} + (\sigma_t^{0})^2/K_t^{ij} \right )
\end{equation}

Direct measurement of the effective width $h$, denoted by \lFPZ{} for the following, of the fracture process zone (FPZ) is another characteristic dimension. We also made this estimation from single tensile tests performed on notched specimen, by treating the fracture energy of each element similarly to acoustic emission \cite{Maji.1998.27,Haidar.2005.201}. A density map of the dissipated elastic energy can be drawn from broken elements. Based on this map, the size of the FPZ can be determined by analyzing the density distribution of dissipated energy around the macrocrack. This distribution, when represented as a probability density function (pdf), can be fitted by a Gaussian distribution in order to extract a width. Rather than that, we choose to rely on the cumulative density function (cdf) of the dissipated energy to determine the size of the FPZ since that curve can be more smoothly defined by sorting the dissipated energy along a direction. The direction chosen here is the one perpendicular to the mean direction of the final crack which may not be strictly perpendicular to the loading direction depending on the microstructure setting. A fit of the cumulated form by a ``Gaussian bell'' allows to assess \lFPZ{} as being four times larger than the standard deviation $\sigma$ of the Gaussian curve. This choice corresponds to a width containing a bit more than 95\% of energy dissipated (provided that only one process zone exists).



\begin{figure}[!htbp]
 \centering
\includegraphics[width=\columnwidth]{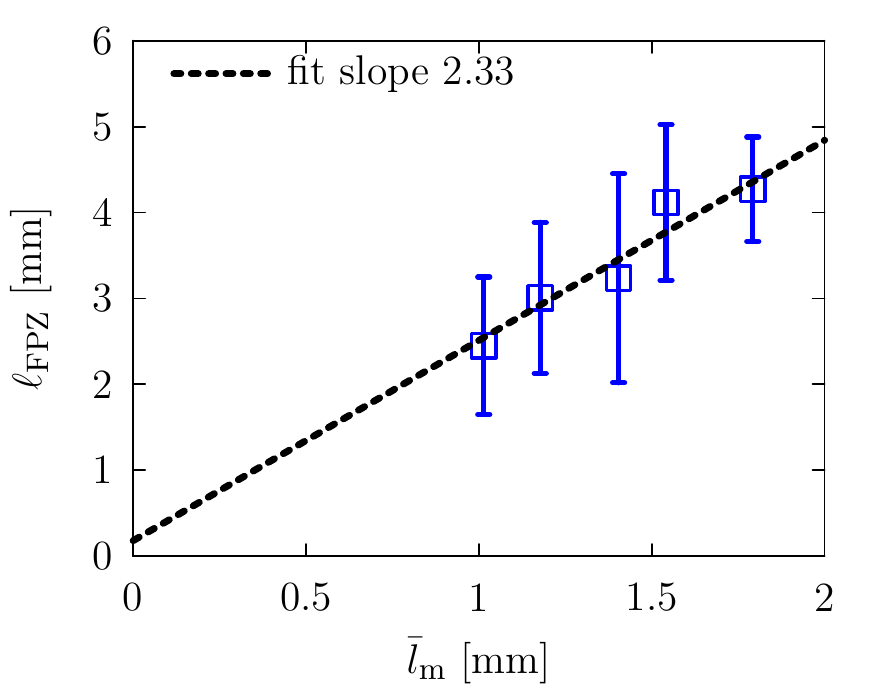}
\caption{Evaluation of the FPZ size with respect to mesh size: the FPZ size \lFPZ{} does statistically vanish under mesh refinement.}
\label{fig:lFPZ_meshSize}
\end{figure}

It is worth pointing out that the FPZ size and the characteristic length of the material determined by lattice simulations also result from the mesh size, \ie{} the lattice element size. This means that the LEM introduces a characteristic length by its mesh. An analysis of the mesh-size influence on the FPZ size is performed. A series of LD tensile tests is performed in which the specimen is discretized with five different mean values of the mesh size $l_{\mathrm{m}}$. Furthermore, for each discretization, five independent meshes are generated by randomly moving the nodes within the radius of $l_{\min}$ (the minimum mesh size) \nomenclature{$l_{\min}$}{the minimum mesh size} to take into consideration mesh orientation effect on \lFPZ. The dependence of the PFZ size on the mesh size is shown in Figure~\ref{fig:lFPZ_meshSize}. $\bar{l}_{\mathrm{m}}$ is the average value of the discretization size \nomenclature{$\bar{l}_{\mathrm{m}}$}{the average value of the discretization size}. As expected, the FPZ size does statistically tend to ``zero'' upon mesh refinement. Note however that the intercept of the fit is not exactly zero, its value is $0.18$~mm. This is probably due to the fact that there are only five discretizations were used and there was not any mesh finer than $1$~mm to be generated for the sake of saving computational time. Once the influence of the mesh on the material internal length is known, it can be subtracted from the relationship between the internal length and the inclusion properties. The latter defines the aim of the present study.

\section{Numerical experiments}

\begin{figure}[!htbp]
 \centering
\includegraphics[width=\columnwidth]{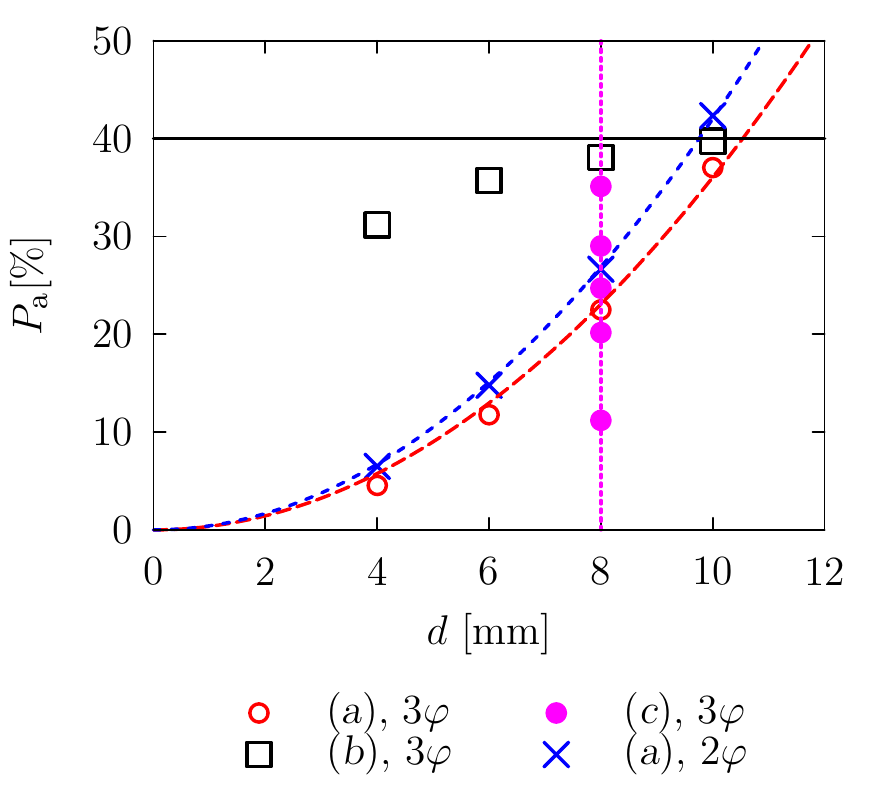}
\caption{Three variation paths (\textit{a}), (\textit{b}) and (\textit{c}) for three-phase material ($3\varphi$) and the variation path (\textit{a}) for two-phase material ($2\varphi$) in the ($P_{\mathrm{a}}$ -- $d$) parameter space for the monodisperse distribution of inclusions.}
\label{fig:variationPath}
\end{figure}

To study the role played by coarse inclusions in the internal length, a number of simulations have been performed. In the modeling of the material mesostructure, inclusions are considered, which are embedded in the matrix separated by the interfacial transition zones (ITZ). The inclusions, matrix and ITZs are assumed to be linear brittle elastic. The inclusions are also assumed to be stiffer and more resistant than the matrix, whereas the ITZs are assumed to be less stiff and with a smaller strength than the matrix. In the following simulations, the stiffness and the strength of inclusions are $10$ times larger than those of the matrix. In turn, the stiffness and the strength of the matrix is $2$ times larger than those of ITZs. Elastic and strength parameters of the matrix are set to values listed in Table~\ref{tab:params}, and they are kept fixed for all simulations.
%
\begin{table*}[ht]
\caption{Elastic and strength parameters used in the bulk of the matrix phase. Corresponding Young's modulus and Poisson's ratio at the macroscopic level are also indicated. \label{tab:params}}
 \centering
\begin{tabular}{l|cccc || cc}
\hline
 Phase $\varphi$  & $\bar{K}_n$  & $\bar{K}_t$ &  $\sigma_n^{0}$  &  $\sigma_t^{0}$  & $E$  &  $\nu$ \\
                  &       (GPa)  &     (GPa)   &     (MPa)    &   (MPa)   &  (GPa)    &  (--) \\
 \hline
 Matrix & 16.50  & 5.10 & 6.07 & 18.21 & 13.20 & 0.20  \\
\hline
\end{tabular}
\end{table*}%
The way coarse inclusions are structured -- referred to as ``grain structure'' in the sequel -- was restricted to two characteristics in this study: the mono-sized grain diameters $d$ and their surface fraction $P_{\mathrm{a}}$ \nomenclature{$d$}{the inclusion diameter} \nomenclature{$P_{\mathrm{a}}$}{surface fraction of inclusions}. In the ($P_{\mathrm{a}}$ -- $d$) parameter space, shown in Figure~\ref{fig:variationPath}, three variation paths were considered: 
\begin{enumerate}
\item[(\textit{a})] varying  $d$ while the positions of inclusions remain the same, $P_{\mathrm{a}}$ thus varies roughly like $d^2$,
\item[(\textit{b})] varying $d$ while $P_{\mathrm{a}}$ is kept at $40$\%\footnote{Note however that the surface fraction of inclusions is not exactly kept constant at $40$\% when changing the inclusion size. This is because of the fact that the smaller the inclusion size, the greater the number of particles are needed, resulting in a greater number of the ITZ elements and consequently leading to a smaller number of inclusion elements.},
\item[(\textit{c})] varying $P_{\mathrm{a}}$ for a given inclusion diameter $d=8$~mm.
\end{enumerate}
%

In addition to the variation of grain structure, the presence of a weak interfacial transition zone between inclusion and matrix phases is analyzed. Without weak ITZ, only two phases ($2\varphi$) are modeled in the sense that the properties of the ITZs defined as in Figure~\ref{fig:phaseDefinition} are those of the matrix. With weak ITZ, a less stiff phase with smaller strength is added in-between inclusions and matrix, bringing the number of phases to three ($3\varphi$).

\begin{figure*}[!htbp]
 \centering
      \begin{subfigure}[b]{0.7\columnwidth}
              \centering
	      \resizebox{1\textwidth}{!}{\includegraphics[width=\textwidth]{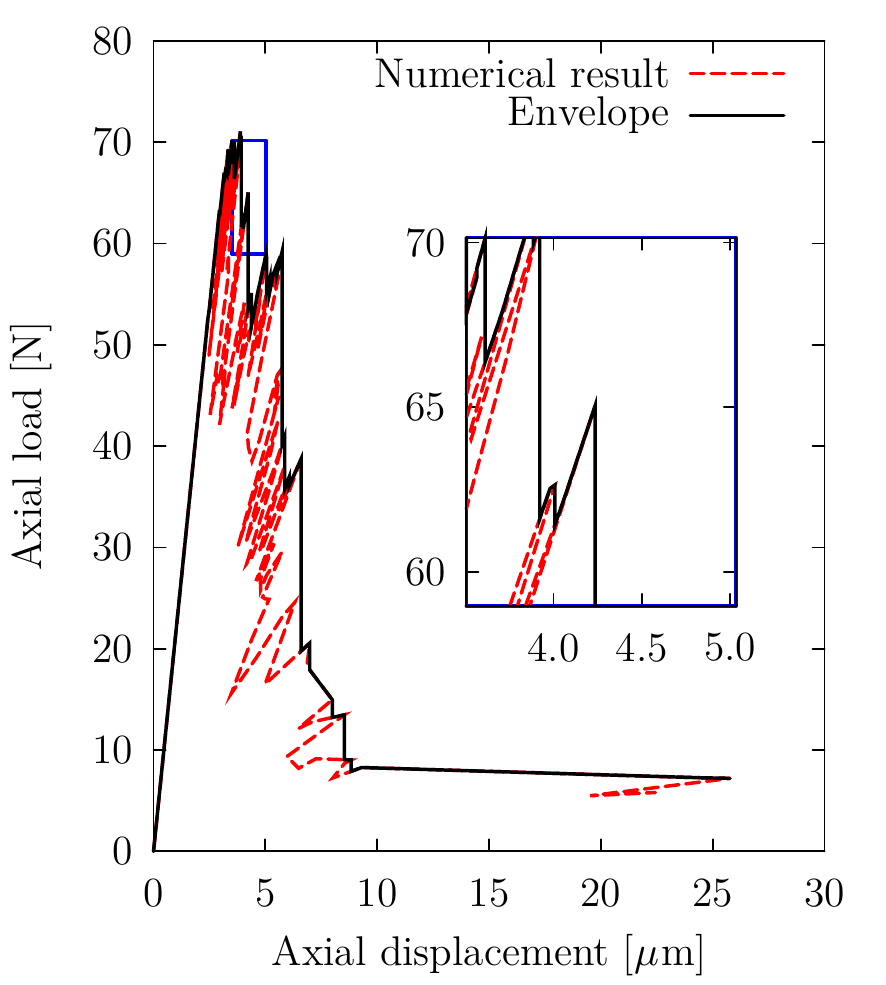}} 
              \caption{}
              \label{fig:force_displacement_TD}
      \end{subfigure}%
      ~ 
      \begin{subfigure}[b]{0.3\columnwidth}
              \centering
              \includegraphics[height=0.3\textheight]{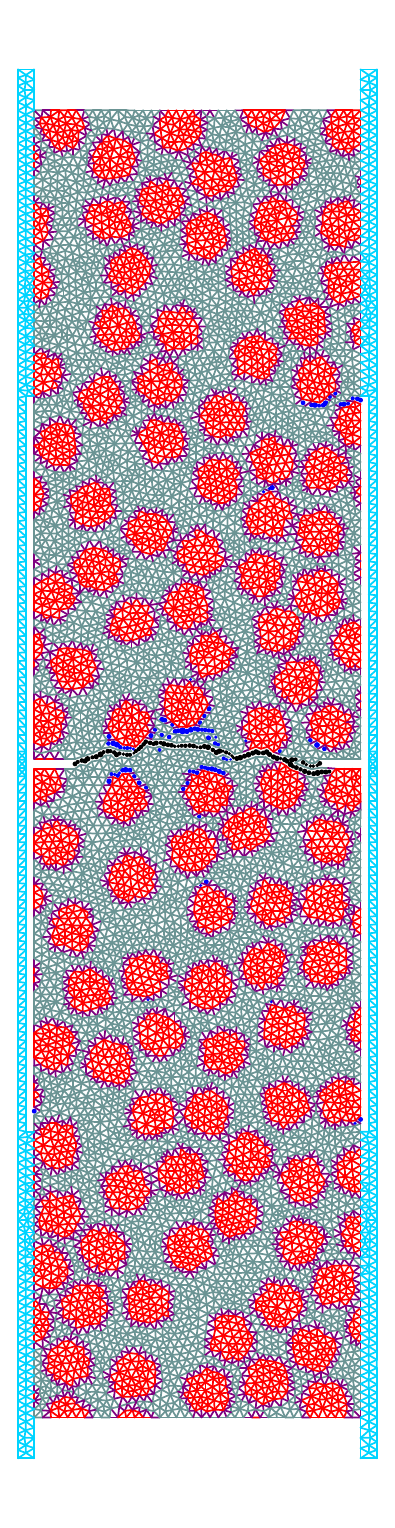}
              \caption{}
              \label{fig:TD_D6P45_NoRandomPos4}
      \end{subfigure}
      ~ 
      \begin{subfigure}[b]{0.7\columnwidth}
              \centering
	      \resizebox{1\textwidth}{!}{\includegraphics[width=\textwidth]{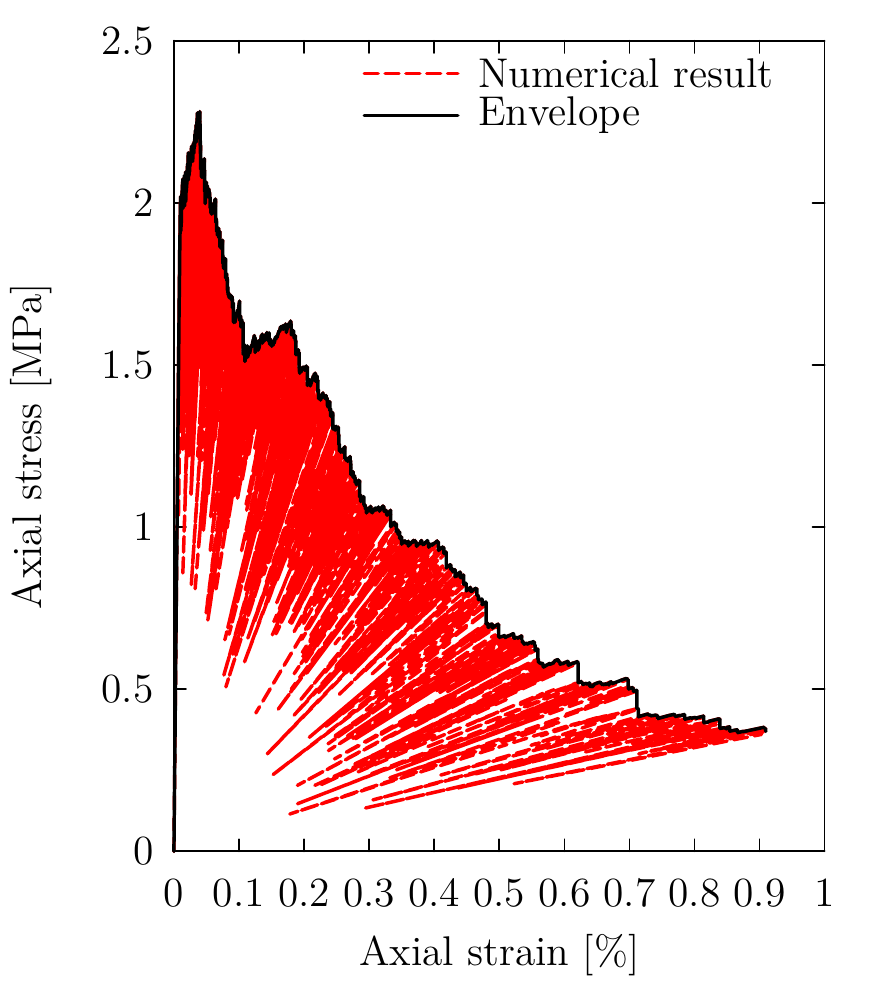}} 
              \caption{}
              \label{fig:stress_strain_PIED}
      \end{subfigure}%
      ~ 
      \begin{subfigure}[b]{0.3\columnwidth}
              \centering
              \includegraphics[height=0.3\textheight]{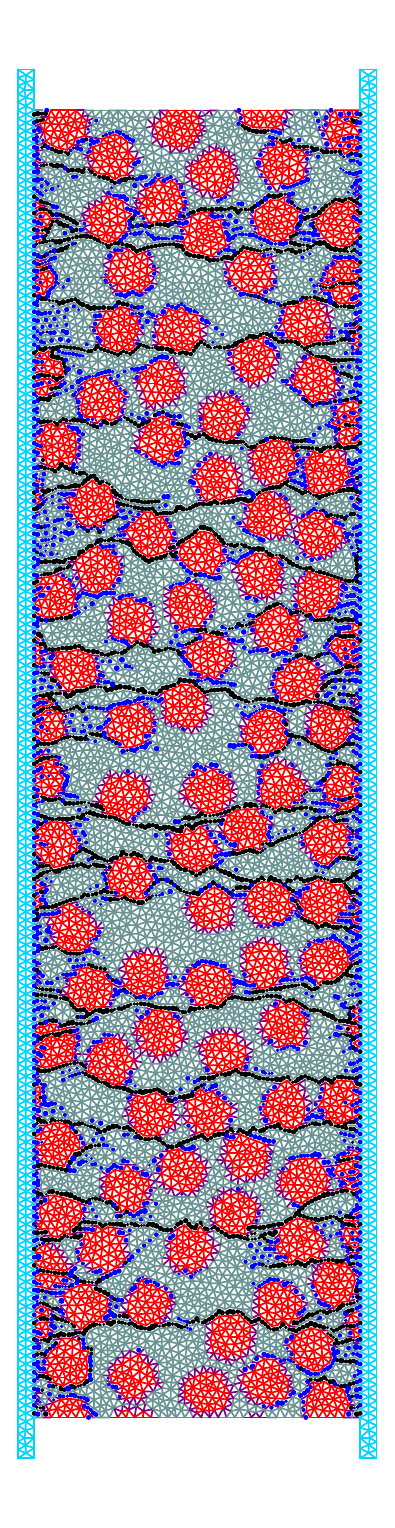}
              \caption{}
              \label{fig:PIED_D6P45_NoRandomPos4}
      \end{subfigure}
\caption{The force-displacement curve (\subref{fig:force_displacement_TD}) and the corresponding crack pattern (\subref{fig:TD_D6P45_NoRandomPos4}) of the localized tensile test on the notched specimen; the stress-strain curve (\subref{fig:stress_strain_PIED}) and the crack pattern (\subref{fig:PIED_D6P45_NoRandomPos4}) of the distributed tensile test on the unnotched specimen.}\label{fig:typical_curves}
\end{figure*}

Typical force-displacement and stress-strain curves obtained for the LD and DD tests, respectively, are shown in Figure~\ref{fig:typical_curves}. The corresponding crack patterns are also presented. It is seen that there is only one macro crack which traverses the notched specimen of the LD test while about fifteen macro cracks cross over the unnotched specimen of the DD test. It shows that the numerical results exhibit disrupted evolution due to the event-driven flow of the simulation. This differs from the experiments in which the displacement is controlled. In fact, the last one is characterized by a monotonic increase of the displacement. Therefore, in order to have a corresponding response, the ``envelope'' of the numerical curve should be taken. The envelope curve is obtained by the so-called smoothing procedure. The procedure is described as follows. By connecting from the first to the last point that describes the specimen state and as soon as a decrease of the displacement is observed, the decrease of the computed load is kept vertically until an intersection with the original curve is observed. The envelope curve then follows the original curve until the new decrease of the displacement is met again and the procedure is repeated. The zoom-in figure in Figure~\ref{fig:force_displacement_TD} shows the procedure. Note that envelope curves were also proposed in~\cite{Arslan1995,Vervuurt1997}. However, when using the envelope curve alone, some essential information may be lost such as a possible snap-back. Also the area under the envelope curve is overestimated. So, the values of $G_f$ and $W_s$ should not be taken from the corresponding areas under the envelopes of the force-displacement and stress-strain curves. Instead, $G_f$ and $W_s$ are directly computed from the stored elastic energies of the broken elements by Equations \eqref{eq:G_f} and \eqref{eq:W_s}.


In all tests presented herein, the characteristic length $\ell_c^0$ intrinsic to the lattice mesh is determined by performing the LD and DD tests on several mesh configurations without inclusions. The intrinsic effective width of the FPZ $\ell_{\mathrm{FPZ}}^0$ is also determined \textit{via} direct measurements. These values are shown in the plots of lengths as if the inclusion diameter or the surface fraction is zero.

\subsection{Key features that may influence the FPZ size}

\subsubsection{Material mesostructure}

\begin{figure}[!htbp]
\begin{center}
\includegraphics[width=\columnwidth]{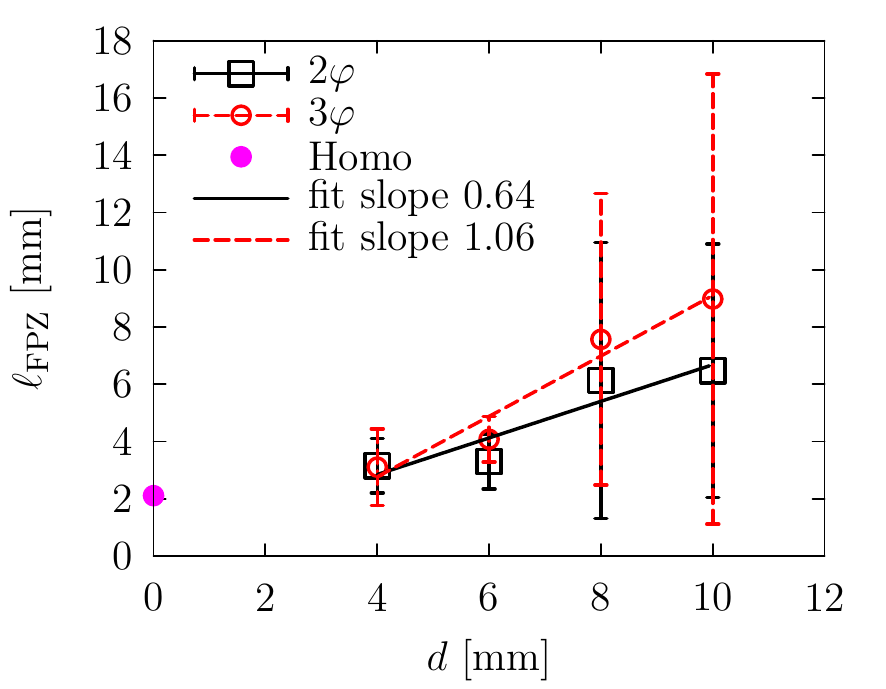}
\caption{Affine relationship between the width of FPZ \lFPZ{} and the diameter $d$ of the inclusions, with ($3\varphi$) or without ($2\varphi$) weak interfacial transition zone between inclusions and matrix is observed using the variation path (\textit{a}).}
\label{fig:l_FPZ_varyPhases}
\end{center}
\end{figure}

\paragraph{Path (\textit{a})}
For concrete materials, it is usual to deem that the characteristic length depends on the aggregate size. Initial investigations with the model have therefore focused on the role of inclusion diameter $d$ on the width \lFPZ{} of the FPZ, while varying $d$ and keeping the positions and the number of inclusions unchanged (variation path (\textit{a})). The evolution of the FPZ size \lFPZ{} with respect to the size of the inclusions $d$ is shown in \Cref{fig:l_FPZ_varyPhases}. In this plot and those that follow, each point with its error bar (standard deviation) requires five measurements and corresponds to the mean value of five values of \lFPZ{} with five independently random distributions of the position of inclusions in the specimen. The lattice mesh used in the simulations provides a width of the FPZ equals to $2.1$~mm. Besides, the best fits of the variation of the mean value of \lFPZ{} with respect to the inclusion size $d$ for the two- and three-phase materials are shown in the figure as well. It is noted that these fits are calculated only from the mean values of \lFPZ{} in the cases of inclusions are introduced, so the value of \lFPZ{} of the homogeneous material is not taken into consideration in the fits. Also, the displayed fitted lines do not necessary mean that an affine relation is enlightened. It must rather be seen as a tendency since the data presents significant variations. As a consequence, the intersection of the fitted line with the vertical axis has no particular meaning \ie, one could also say the fit is only valid between the limits studied.

\begin{figure}[!htbp]
 \centering
\includegraphics[width=\columnwidth]{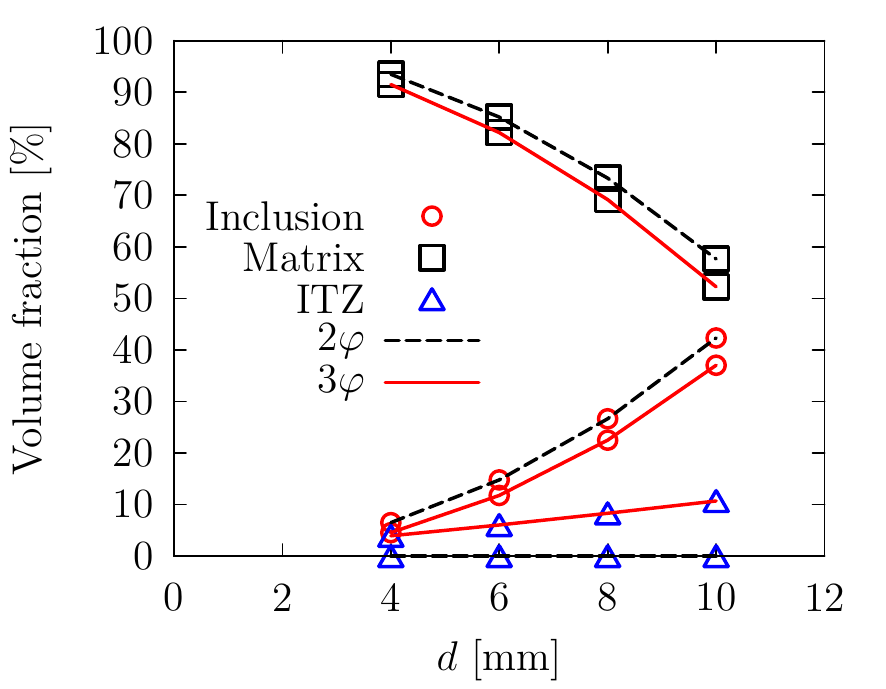}
\caption{Evaluation of surface fraction of each phase.}
\label{fig:plotPhasesFraction}
\end{figure}

The main observation from the \Cref{fig:l_FPZ_varyPhases} is that when the inclusions are introduced, they have a strong effect on the FPZ size in both two- and three-phase materials. First, the mean values of \lFPZ{} in the case of heterogeneous material are greater then the value of \lFPZ{} in the case of homogeneous one. Second, in the case of inclusions are introduced, the fitted slope of the mean values of \lFPZ{} of the three-phase material is greater than that of the two-phase material. This means that when the ITZ is taken into account, the inclusion size plays a stronger effect on the variation of \lFPZ{} than the case in which the ITZ is not taken into consideration. So, according to our model, the internal length does not only depend on the size of the inclusions but also their constituents and therefore the presence of ITZ. The second observation is probably explained by the increase of the ITZ fraction when increasing the inclusion size of the three-phase material, see \Cref{fig:plotPhasesFraction}. Here, the ITZ plays a role of attractive zones for the crack propagation because of their lower strengths and stiffnesses. Accordingly, the greater fraction of the ITZ results in the larger mean value of \lFPZ{} compared to the mean value of \lFPZ{} of the two-phase material (in which the ITZ fraction is zero). In the case of $d=4$~mm, the mean value of \lFPZ{} of the three-phase material does not differ from that of the two-phase one. This is related to the fact that the matrix prevails in the mesostructure in the case of $d=4$~mm, as shown in~\Cref{fig:plotPhasesFraction}, and thus few inclusions are found on the crack path.

Furthermore, bigger values of the standard deviations are observed when increasing the particle size of the three-phase material as well as of the two-phase one even though this is less clearly observed in the two-phase material than the three-phase one. We believe that increasing the inclusion size results in the increase of the inclusion fraction and as a result, the spatial distribution of inclusions plays a stronger role in the resulting value of \lFPZ. In two-phase material, by comparison between $d=8$~mm and $d=10$~mm, the standard deviation of \lFPZ{} does not significantly change. This is due to the fact that from $d=8$~mm, the inclusions get dense in the mesostructure of the material, in that a change of the position of the inclusions does not have a strong effect on the value of \lFPZ. However, in three-phase material, the spatial distribution of inclusions still make sense on the variation of \lFPZ{} even though the inclusions get dense. This is reflected by the greater value of the standard deviation of \lFPZ{} in the case of $d=10$~mm than that in the case of $d=8$~mm, of the three-phase material. There is no doubt that this is due to the effect of the ITZ.

\paragraph{Path (\textit{b})}
A second series of tests is performed by following the variation path (\textit{b}), that is with fixed surface fraction of inclusions and varied inclusion diameter. The reason of this choice relies on the fact that the fundamental role of inclusion size $d$ must be checked while keeping other parameters unchanged to suppress their possible effect. 

\begin{figure}[!htbp]
 \centering
\includegraphics[width=\columnwidth]{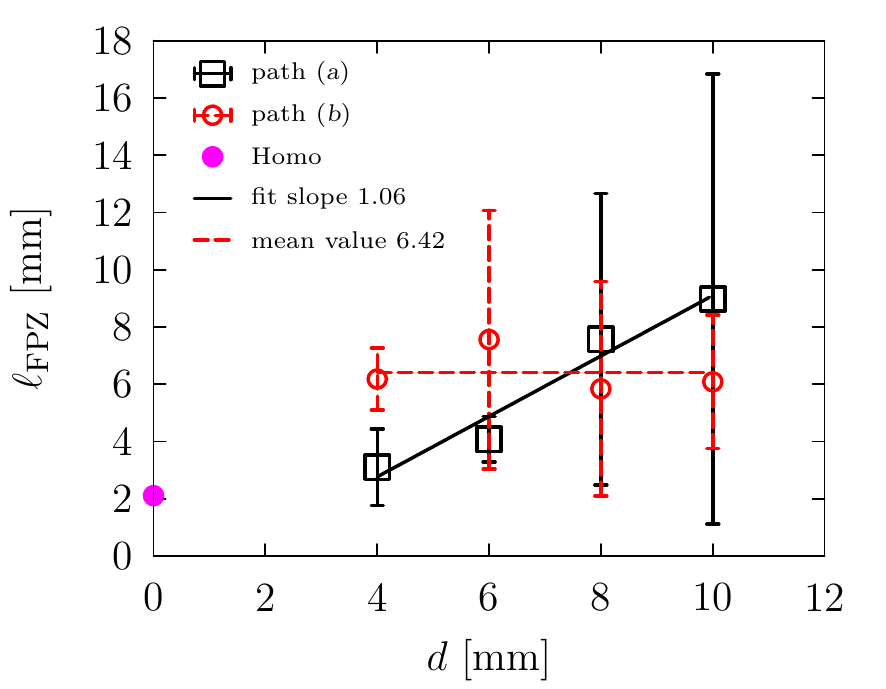}
\caption{Relation between the FPZ size \lFPZ{} and inclusion size $d$ with respect to the variation paths (\textit{a}) and (\textit{b}).}
\label{fig:lFPZ_N55vsP45}
\end{figure}

\Cref{fig:lFPZ_N55vsP45} shows the plot of the mean value of the FPZ size \lFPZ{} with respect to the size $d$ of inclusions of the path (\textit{b}) of variation. For the sake of comparison, the same plot of the path (\textit{a}) is shown as well. Surprisingly, it exhibits that the mean value of the FPZ size does not depend on the inclusion size for the path (\textit{b}) of variation. It means that the FPZ size developed in this type of model material (brittle elastic) may not always be related to the inclusion size itself as usually observed in the literature. The observation is in agreement with that of~\cite{Skar.zynski2011}, in which the width of the FPZ was experimentally measured on the surface of concrete specimens using a Digital Image Correlation (DIC) technique. However, it is in contrast to the results of~\cite{Mihashi1996,Otsuka2000} for concrete material, in which the experiments were carried out with X-rays and three-dimensional Acoustic Emission techniques leading to the conclusion that the width of the FPZ increases with the increase of the maximum inclusion size. But for the path (\textit{a}) of variation, as previously shown, it is seen that the mean values of \lFPZ{} increases with the increase of inclusion size $d$ that also results in the increase of the inclusion surface fraction.

\begin{figure*}[!htbp]
\centering
      \begin{subfigure}[b]{0.5\columnwidth}
              \centering
              \includegraphics[trim = 1mm 0mm 1mm 0mm, clip = true,width=\textwidth]{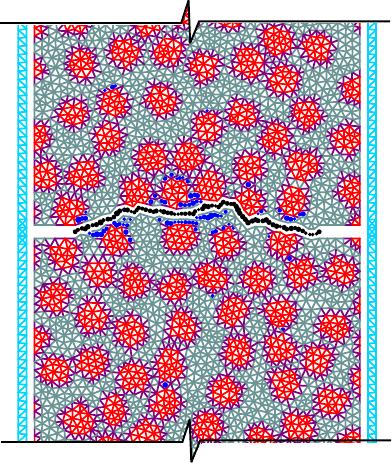}
              \caption{$d=4$, \lFPZ{} $=5.47$}
              \label{fig:D4P45_Pos1}
      \end{subfigure}%
      ~ 
      \begin{subfigure}[b]{0.5\columnwidth}
              \centering
              \includegraphics[trim = 1mm 0mm 1mm 0mm, clip = true,width=\textwidth]{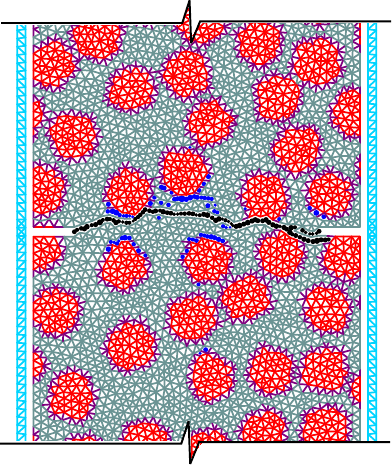}
              \caption{$d=6$, \lFPZ{} $=4.94$}
              \label{fig:D6P45_Pos4}
      \end{subfigure}%
      ~ 
      \begin{subfigure}[b]{0.5\columnwidth}
              \centering
              \includegraphics[trim = 1mm 0mm 1mm 0mm, clip = true,width=\textwidth]{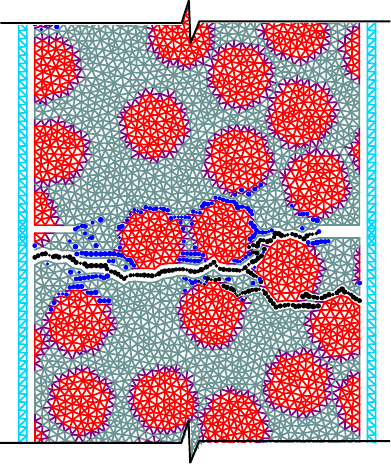}
              \caption{$d=8$, \lFPZ{} $=12.46$}
              \label{fig:D8P45_Pos3}
      \end{subfigure}%
      ~ 
      \begin{subfigure}[b]{0.5\columnwidth}
              \centering
              \includegraphics[trim = 1mm 0mm 1mm 0mm, clip = true,width=\textwidth]{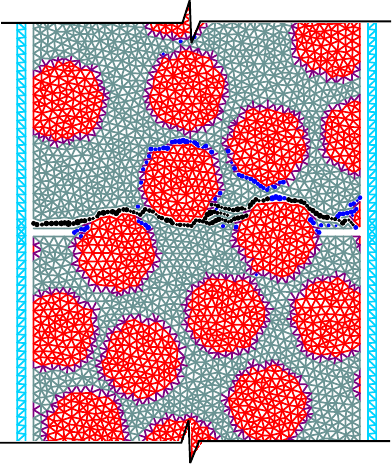}
              \caption{$d=10$, \lFPZ{} $=4.75$}
              \label{fig:D10P45_Pos3}
      \end{subfigure}%
\caption{Crack patterns and the corresponding FPZ size \lFPZ~[mm] of the LD tensile tests on the specimen made of the material with different inclusion size $d$~[mm] of the variation path (\textit{b}).}\label{fig:3Phases_P45}
\end{figure*}

\begin{figure*}[!htbp]
\centering
      \begin{subfigure}[b]{0.5\columnwidth}
              \centering
              \includegraphics[trim = 1mm 0mm 1mm 0mm, clip = true,width=\textwidth]{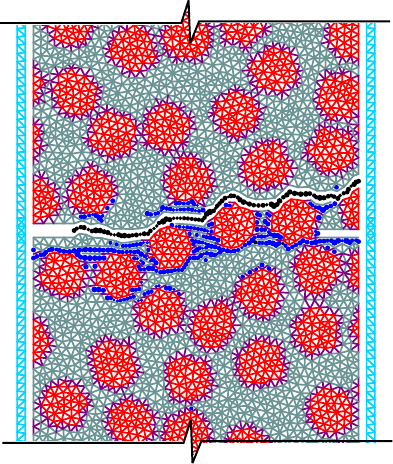}
              \caption{\lFPZ{} $=10.60$}
              \label{fig:D6P45_Pos1}
      \end{subfigure}%
      ~ 
      \begin{subfigure}[b]{0.5\columnwidth}
              \centering
              \includegraphics[trim = 1mm 0mm 1mm 0mm, clip = true,width=\textwidth]{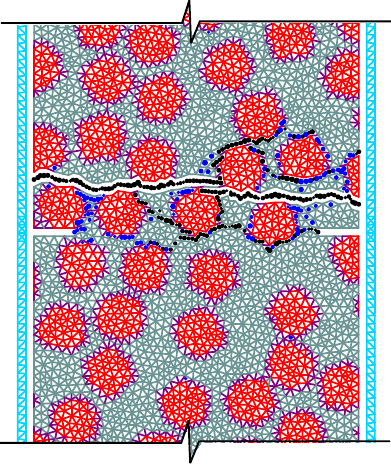}
              \caption{\lFPZ{} $=14.01$}
              \label{fig:D6P45_Pos2}
      \end{subfigure}%
      ~ 
      \begin{subfigure}[b]{0.5\columnwidth}
              \centering
              \includegraphics[trim = 1mm 0mm 1mm 0mm, clip = true,width=\textwidth]{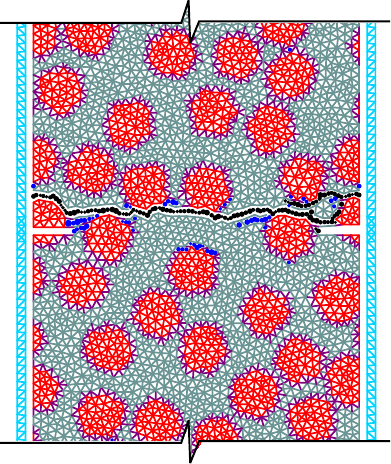}
              \caption{\lFPZ{} $=3.81$}
              \label{fig:D6P45_Pos3}
      \end{subfigure}%
      ~ 
      \begin{subfigure}[b]{0.5\columnwidth}
              \centering
              \includegraphics[trim = 1mm 0mm 1mm 0mm, clip = true,width=\textwidth]{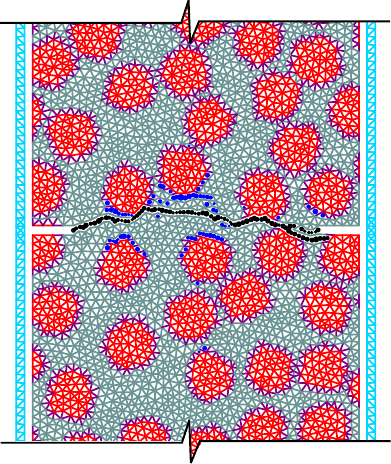}
              \caption{\lFPZ{} $=4.94$}
              \label{fig:D6P45_Pos4_}
      \end{subfigure}%
\caption{Crack patterns and the corresponding FPZ size \lFPZ~[mm] of the LD tensile tests on the specimen made of the material with inclusion size $d=6$~mm with four different distributions of the inclusions.}\label{fig:3Phases_P45_d6}
\end{figure*}

\Cref{fig:3Phases_P45} shows the crack patterns and the corresponding value of \lFPZ{} obtained when changing the inclusion size $d$ within the variation path (\textit{b}). The figure shows for each inclusion size $d$ only one random distribution of inclusions in the specimen. It seems that the position of inclusions around the notches have an essential role on the resulting FPZ size. In fact, the crack is always initiated at the weak ITZ between the inclusion and matrix phases. With regard to the position of a notch -- that can also be seen as a ``weak link'' -- the crack is then propagated \textit{via} the development of microcracks and at the end the macrocrack is formed by connecting the notch(es) and the broken elements (mainly in the ITZs). However, sometimes an inclusion is found just in front of the notch(es) and it plays a role of an obstacle that prevents the rupture of elements in the vicinity of the notch(es) and consequently, prevents the macrocrack to reach the notch(es). In this case, the macrocrack is finally formed by mainly connecting the broken ITZ elements. Therefore, the spatial distribution (positions) of the inclusions actually have an important role on the FPZ size in conjunction with the size of the inclusions. Nevertheless, for the case where the reference surface fraction of inclusion is kept almost constant (path (\textit{b})), the spacing between the inclusions seems to be constant regardless of the size of the particles, and thus the spatial distribution of the particles prevails more and more on their size in the resulting FPZ size \lFPZ. Actually, as shown in~\Cref{fig:3Phases_P45_d6} in which four different sets of inclusion positions with the diameter being $6$~mm,
the values of \lFPZ{} are finally different depending on the spatial distribution of the inclusions with regard to the notch position. This explains why changing the size of the inclusions according to the path (\textit{b}) does not change the value of \lFPZ{} averaged over five random spatial distributions of inclusions. On the contrary, within the path (\textit{a}) of variation, changing the inclusion size leads to a change in the inclusion surface fraction together with the spacing between inclusions, and the FPZ size is affected by not only the size of inclusions but also the other structuring parameters (position and surface fraction of inclusions in our case). Still, the smaller the inclusion particle size, the larger the spacing between the particles in the path (\textit{a}). This leads to the weaker influence of the spatial distribution of inclusions observed on the FPZ size. It is revealed in \Cref{fig:lFPZ_N55vsP45} by the value of the standard deviation that is increased with the inclusion size.

\begin{figure}[!htbp]
 \centering
\includegraphics[width=\columnwidth]{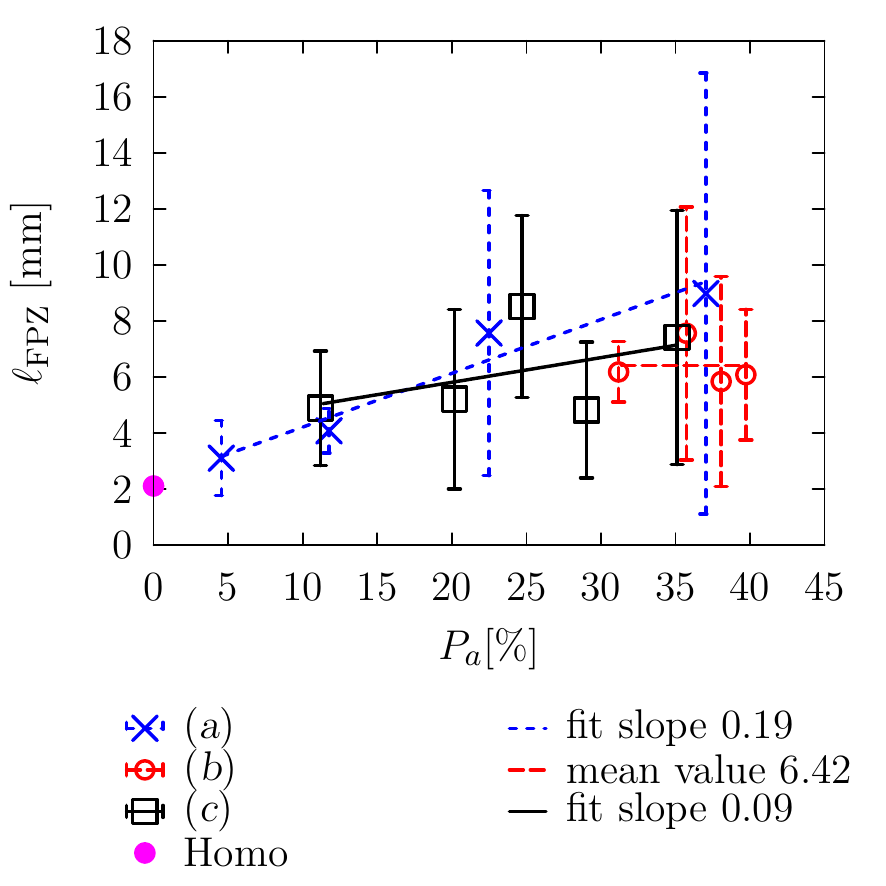}
\caption{Variation of \lFPZ{} according to the inclusion surface fraction $P_{\mathrm{a}}$ of the three variation paths (\textit{a}), (\textit{b}) and (\textit{c}).}
\label{fig:lFPZ_P45vsN55vsD8varyGrainDensity}
\end{figure}

\paragraph{Path (\textit{c})}
A third series of tests were performed by following the variation path (\textit{c}) in which the inclusion surface fraction is varied while keeping the inclusion size constant at $8$~mm in order to evaluate the only influence of the inclusion surface fraction (or equivalently the inclusion spacing, since the latter is inversely proportional to the former) on the FPZ size \lFPZ. \Cref{fig:lFPZ_P45vsN55vsD8varyGrainDensity} shows the variation of the mean value of \lFPZ{} with respect to the inclusion surface fraction $P_{\mathrm{a}}$. For the sake of comparison, the results of above studies for the variation paths (\textit{a}) and (\textit{b}) are plotted as well, but in the (\lFPZ--$P_{\mathrm{a}}$) space. The main observation is that the mean value of \lFPZ{} of the path (\textit{a}) and the path (\textit{c}) does increase with the increase of the inclusion surface fraction $P_{\mathrm{a}}$, whereas that of the path (\textit{b}) does not change. This is simply explained by the fact that the spacing between inclusions decreases with the increase of the inclusion surface fraction within the variation paths (\textit{a}) and (\textit{c}), whereas it seems to be ``constant'' (or hardly changed) within the path (\textit{b}). By comparison the path (\textit{c}) with the path (\textit{a}), it is observed, however, that the increase rate of \lFPZ{} with respect to $P_{\mathrm{a}}$, which is represented by the fitted slope, of the path (\textit{c}) is smaller than that of the path (\textit{a}). A suitable explanation for this observation is that within the path (\textit{a}), the size of and the spacing between the inclusions do change (increase and decrease, respectively) at the same time with respect to the increase of $P_{\mathrm{a}}$ whereas only the spacing of the inclusions does decrease with respect to the increase of $P_{\mathrm{a}}$ within the path (\textit{c}). \emph{Therefore, the observation could lead to the evidence that the FPZ size depends on both the inclusion spacing (which is just a consequence of the inclusion surface fraction) and inclusion size}.

\begin{figure}[!htbp]
 \centering
\includegraphics[width=0.8\columnwidth]{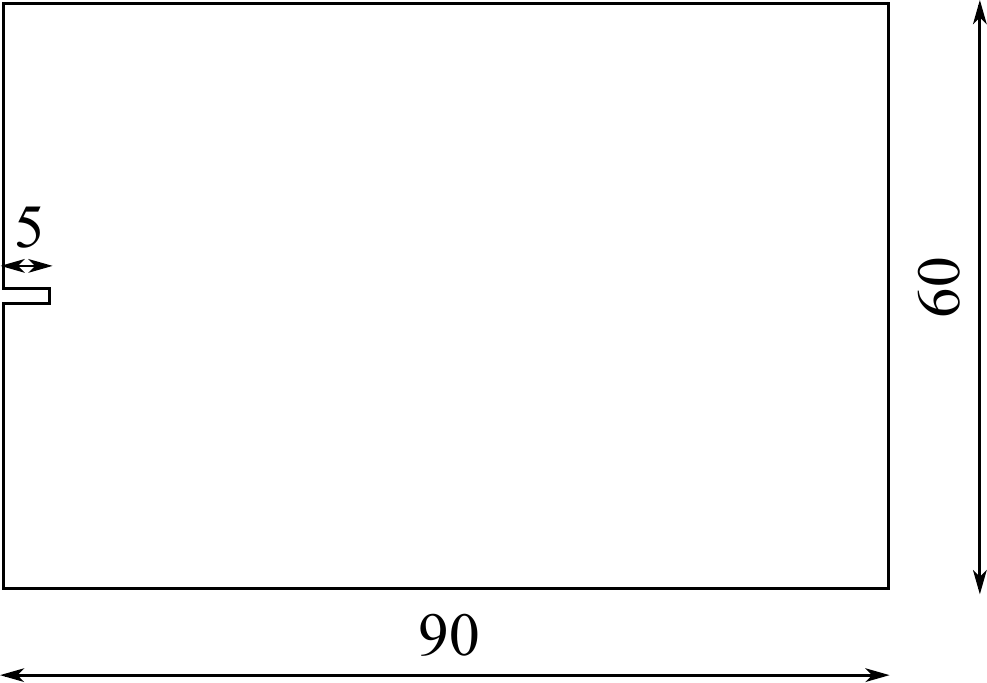}
\caption{Specimen dimensions [mm].}
\label{fig:specimen90x60}
\end{figure}

\paragraph{Larger specimen width compared to inclusion size}
The above studies of \lFPZ{} are performed on slender specimens (small ligament size, \ie{} in the order of $3 \times d$ compared to inclusion size $d$). These slender specimens were used to ensure the damage distribution in DD-tests as homogeneous as possible for studies of the characteristic length \lc{} which is presented in \Cref{sec:lc_vs_lFPZ}. However, it was shown that the FPZ size \lFPZ{} performed on these slender specimens strongly depends on the position of inclusions. This can be considered as a drawback for an attempt to correlate the FPZ size and the inclusion size. Therefore, it would be better if the study of \lFPZ{} is performed on a larger specimen size compared to the inclusion size. To this end, tensile tests are performed on the specimen shown in \Cref{fig:specimen90x60} with mono-sized inclusion structures and the inclusion size is varied by taking the value $4$, $6$, $8$ and $10$~mm. Three variation paths above are also considered here.

\begin{figure}[!htbp]
 \centering
\includegraphics[width=\columnwidth]{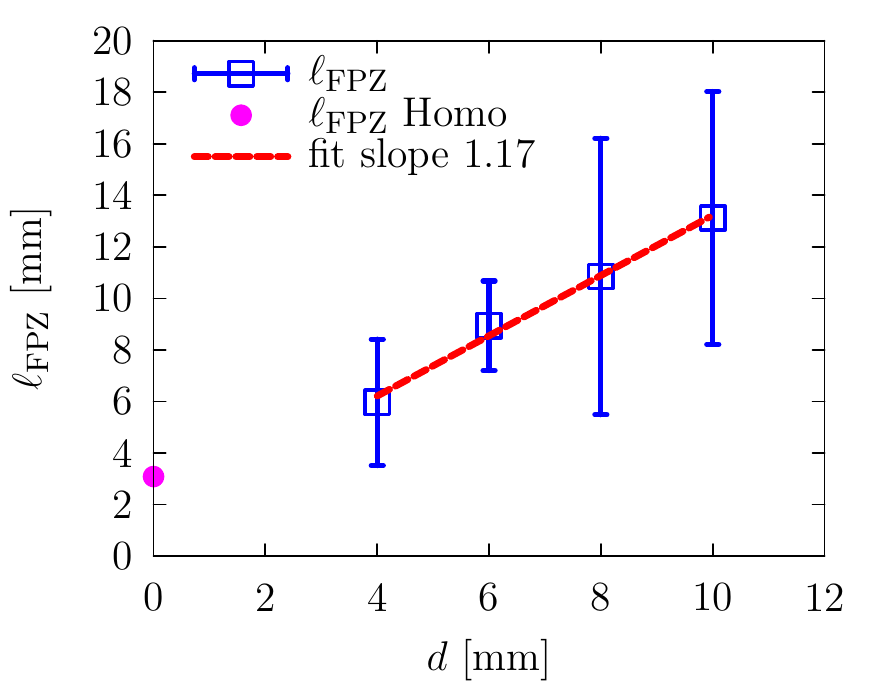}
\caption{Variation of \lFPZ{} according to the inclusion size $d$ of the variation path (\textit{a}).}
\label{fig:lFPZ_N45_specimen90x60}
\end{figure}

\begin{figure}[!htbp]
 \centering
\includegraphics[width=\columnwidth]{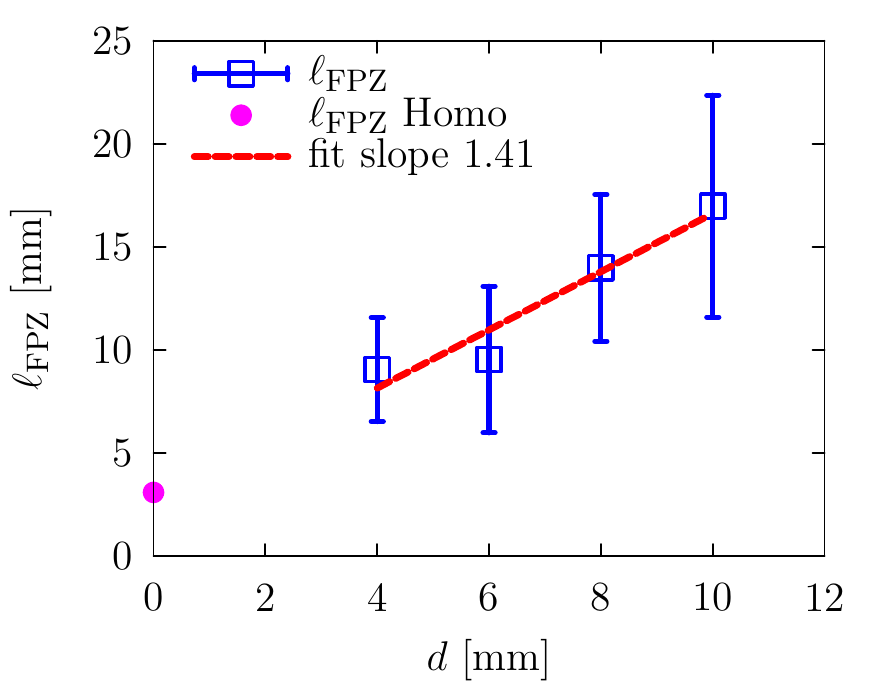}
\caption{Variation of \lFPZ{} according to the inclusion size $d$ of the variation path (\textit{b}).}
\label{fig:lFPZ_P45_specimen90x60}
\end{figure}

\begin{figure}[!htbp]
 \centering
\includegraphics[width=\columnwidth]{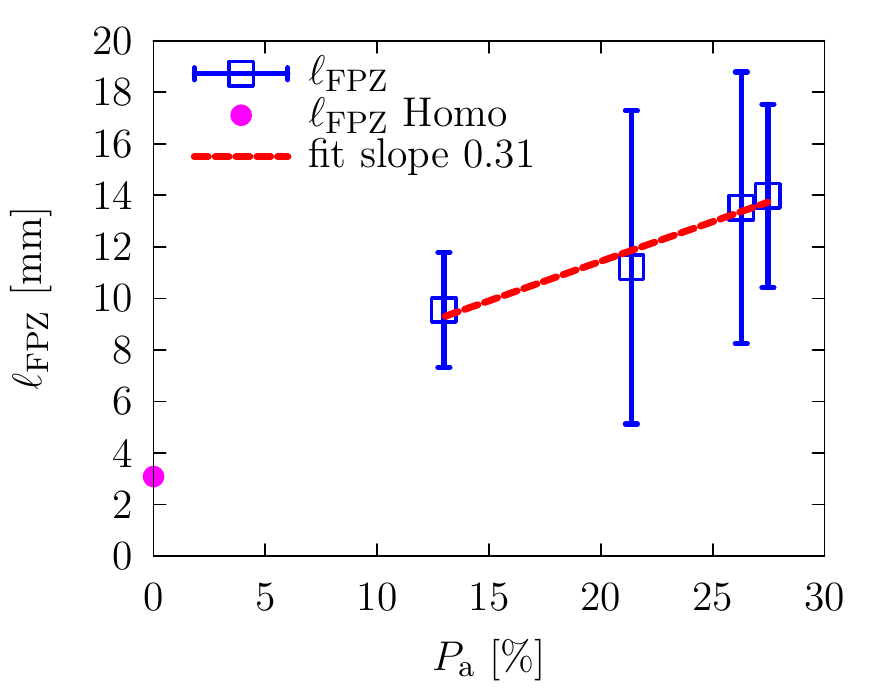}
\caption{Variation of \lFPZ{} according to the inclusion surface fraction $P_{\mathrm{a}}$ of the variation path (\textit{c}).}
\label{fig:lFPZ_D8varyP_specimen90x60}
\end{figure}

\Cref{fig:lFPZ_N45_specimen90x60,fig:lFPZ_P45_specimen90x60} show the variations of \lFPZ{} with respect to the inclusion size $d$ when keeping the number and the position of inclusions unchanged (path (\textit{a})) and when keeping the surface fraction of inclusion constant at $45$\% (path (\textit{b})), respectively. \Cref{fig:lFPZ_D8varyP_specimen90x60} shows the variation of \lFPZ{} with respect to the surface fraction $P_{\mathrm{a}}$ when keeping the inclusion size constant at $8$~mm (path (\textit{c})). It exhibits that the variation of \lFPZ{} with respect to the position of inclusions is less important than the previous cases. Also, better fit is obtained with higher coefficients of correlation ($0.99$, $0.93$ and $0.96$ for paths (\textit{a}), (\textit{b}) and (\textit{c}) respectively).

By comparison between \Cref{fig:lFPZ_N45_specimen90x60,fig:lFPZ_P45_specimen90x60}, it can be seen that, in contrast to the results tested on the slender specimens, a higher influence of the inclusion size $d$ on the FPZ size \lFPZ{} of the variation path (\textit{b}) compared to that of the variation path (\textit{a}). Indeed, a higher value of the fit slope is obtained within the variation path (\textit{b}). It is likely due to the fact that when analyzing on the larger specimen (compared to the inclusion size), the sensitivity of \lFPZ{} with respect to the position of inclusions is less important than testing on the slender specimen, and thus the role of the inclusion size in the FPZ size prevails over the position. So, in conjunction with the influence of the inclusion surface fraction on the FPZ size (which can be observed in \Cref{fig:lFPZ_D8varyP_specimen90x60}), the higher influence of the inclusion size on the FPZ size is obtained within the path (\textit{b}) than that within the path (\textit{a}) because varying the inclusion size in the path (\textit{b}) is combined with a higher surface fraction of inclusions than in the path (\textit{a}).

Therefore, a partial conclusion is that depending on the relative size between the macroscopic size (specimen size) and the mesoscopic size (inclusion size), the influence of the mesostructure on the FPZ size is different. When this relative size is small, the effect of the position of inclusions or the spacing between inclusions of the mesostructure prevails over the effect of the inclusion surface fraction. When the relative size is more important ($\geq 5$), an inverse influence is observed. In any case, the influence of the inclusion size on the FPZ size is always recognized.

\subsubsection{Ligament size}

\begin{figure}[!htbp]
 \centering
 \includegraphics[width=0.7\columnwidth]{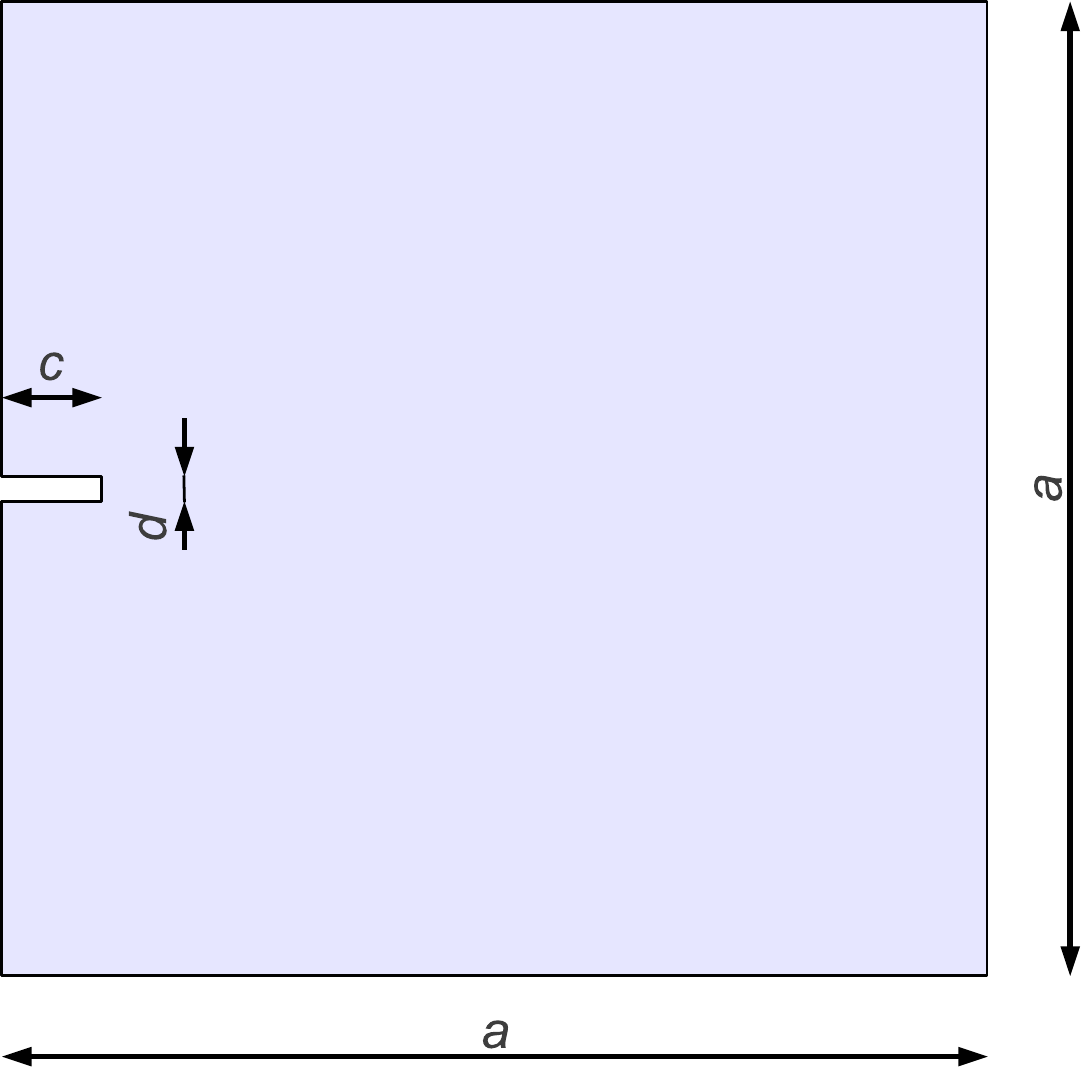}
 \caption{Specimen geometry.}
 \label{fig:ligament_Specimen}
\end{figure}

The specimen geometry is shown in \Cref{fig:ligament_Specimen} and the dimensions of the specimens used in the numerical test are given in \Cref{tab:ligament_specimenDimensions}, with the same size but with different notch lengths that results in different ligament lengths: $90$, $80$, $65$ and $50$~mm. They are labeled by L, M, S and XS respectively for convenience.

\begin{table*}[!htbp]
\caption{Specimen dimensions~[mm].}
 \centering
\begin{tabular}{l l l l l}
\hline
 Ligament size & Long (L) & Medium (M) & Small (S) & eXtra Small (XS) \\
\hline
Specimen size: $a$ & 100 & 100 & 100 & 100 \\
Notch length: $c$ & 10 & 20 & 35 & 50 \\
Notch width: $d$ & 2 & 2 & 2 & 2 \\
Ligament length: $a-c$ & 90 & 80 & 65 & 50 \\
\hline
\end{tabular}
\label{tab:ligament_specimenDimensions}
\end{table*}

In order to study the influence of the maximum inclusion size $d_{\max}$ on the FPZ size, the tensile tests are performed on the specimens made of the material with a polydisperse inclusion structure with $d_{\max}$ being $6.3$, $8$, $10$, $12.5$ and $16$~mm and the reference inclusion surface fraction is kept constant at $45$\%. The minimum inclusion size $d_{\min}$ is $3.15$~mm. All the inclusion gradings are generated by the Fuller's curve which is an ``ideal`` grading curve~\cite{Fuller.1906.67}. In the study, for each inclusion grading up to $d_{\max}$, five inclusion structure realizations are generated with independently random distribution of inclusion positions. The specimens are loaded in tension by directly imposing the vertical displacement increment on the nodes of the top boundary of the specimens while vertically fixing the nodes of their bottom boundary.

\begin{figure}[!htbp]
\centering
  \includegraphics[width=\columnwidth]{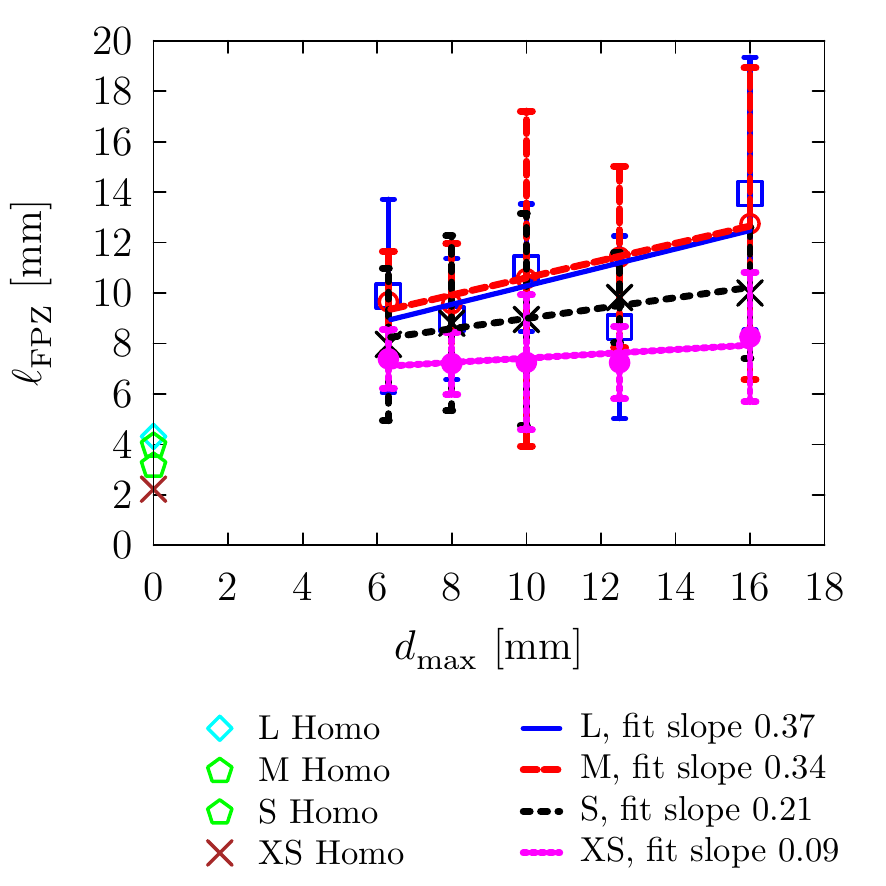}
  \caption{Influence of the ligament length on the variation of the FPZ size \lFPZ{} with respect to the maximum inclusion size $d_{\max}$: L (long ligament), M (medium ligament), S (small ligament), XS (extra small ligament).}
\label{fig:lFPZ_ligamentSize}
\end{figure}

\Cref{fig:lFPZ_ligamentSize} shows the relationship between the FPZ size \lFPZ{} with respect to the maximum inclusion size $d_{\max}$ for the specimens corresponding to four ligament lengths L, M, S and XS. This figure also shows the FPZ size of L, M, S, XS specimens in which no inclusion structure has been introduced. It is seen that, when the inclusion structures are introduced, it always results in a larger FPZ size than the one computed with the homogeneous cases. For a given value of $d_{\max}$, the mean value of \lFPZ{} is systematically increased when the ligament size is increased. In addition, the increase rate of \lFPZ{} is also increased with $d_{\max}$. It results in an higher slope of variation of \lFPZ{} as a function of $d_{\max}$ for a larger ligament size. It is also observed that the increase rate of the slope of variation of \lFPZ{} decreases with the increase of the ligament size from XS specimens to L specimens. So, a stabilized value of variation slope can be achieved for specimens with the ligament size being in order of specimen width. When the ligament size is half (and may be lower by extrapolation) specimen width (the XS specimens), the variation slope is negligible, which means that the inclusion size appears to have no influence on the mean value of \lFPZ. It may suggest that the FPZ has not enough time to develop completely within the specimens with ``too short'' ligament length. Between these limits, the slope variation evolves progressively, indicating that both the inclusion structure and the specimen dimension itself can play a role on the FPZ size. The maximum-inclusion-size independence of \lFPZ{} for the specimens with too short ligament length is in agreement with the previous study performed on the specimen which also has a short ligament length.



\begin{figure}[!htbp]
\centering
      \begin{subfigure}[b]{0.5\columnwidth}
              \caption*{XS}
              \centering
              \includegraphics[trim = 0mm 0mm 0mm 0mm, clip = true,width=\textwidth]{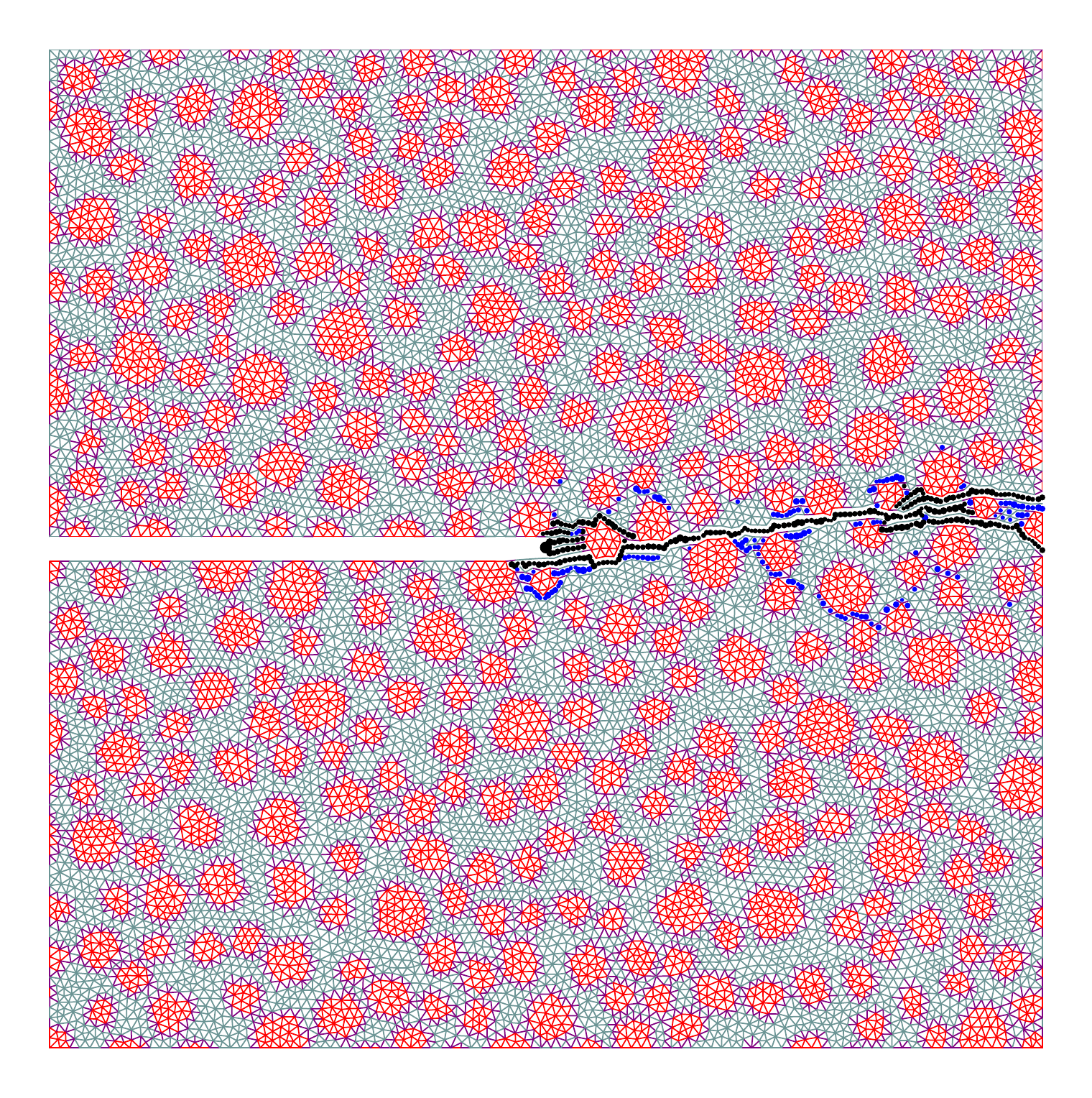}
              \caption{$d_{\max}=6.3$, \lFPZ{} $=7.97$}
              \label{fig:XS_Dmax6.3_Pos4}
      \end{subfigure}%
      ~ 
      \begin{subfigure}[b]{0.5\columnwidth}
              \caption*{L}
              \centering
              \includegraphics[trim = 0mm 0mm 0mm 0mm, clip = true,width=\textwidth]{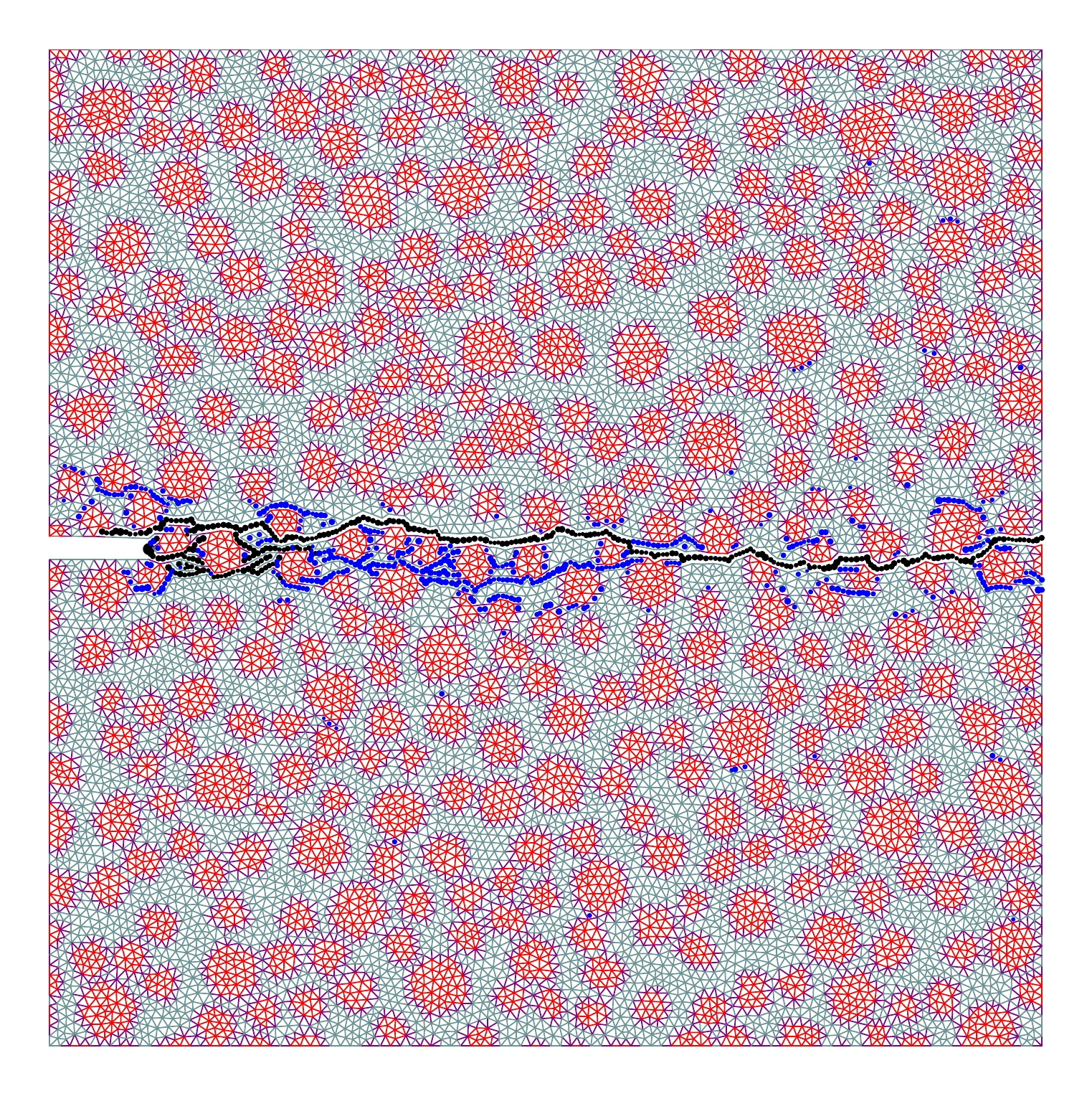}
              \caption{$d_{\max}=6.3$, \lFPZ{} $=8.57$}
              \label{fig:L_Dmax6.3_Pos1}
      \end{subfigure}%

      \begin{subfigure}[b]{0.5\columnwidth}
              \centering
              \includegraphics[trim = 0mm 0mm 0mm 0mm, clip = true,width=\textwidth]{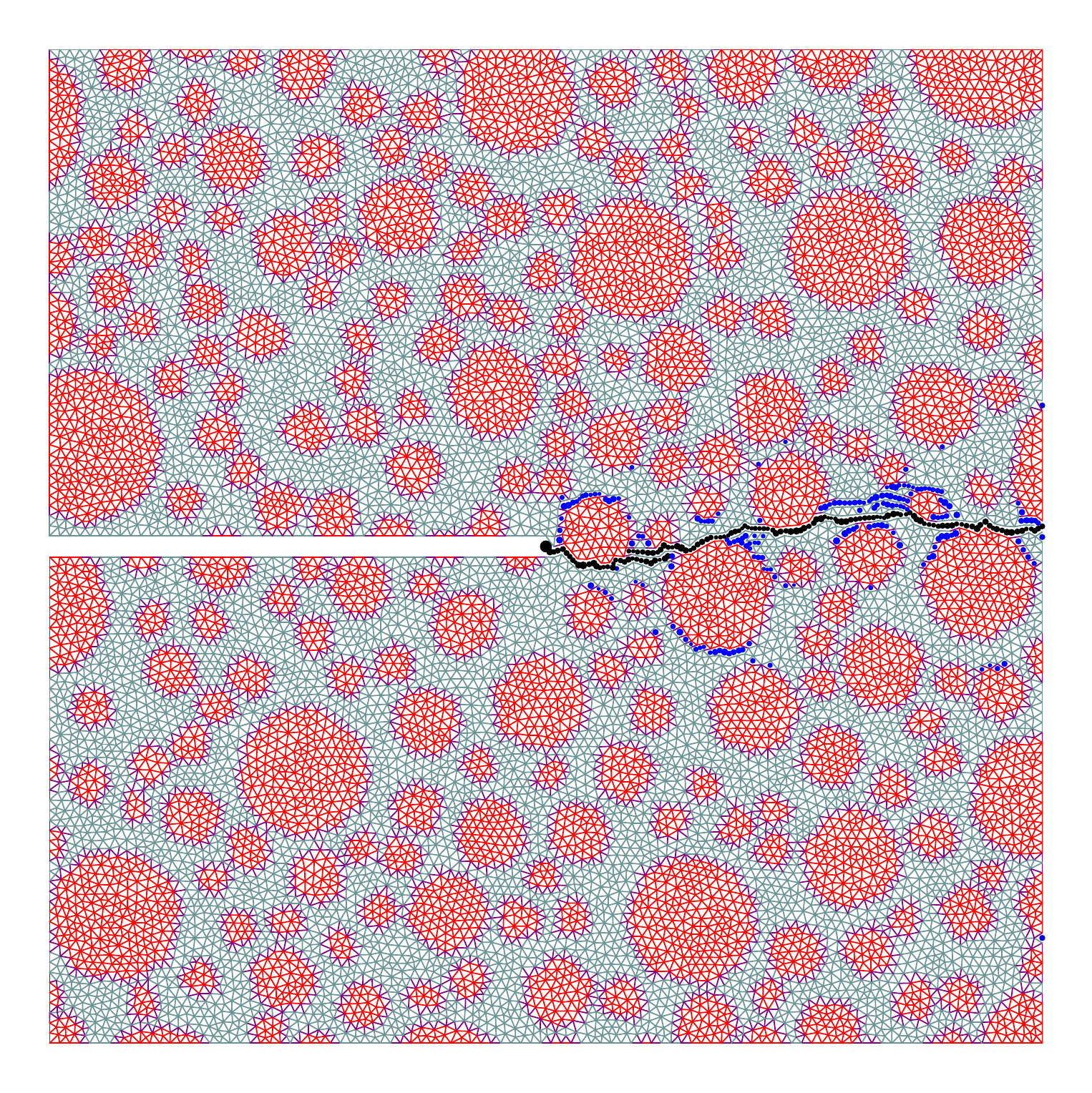}
              \caption{$d_{\max}=16$, \lFPZ{} $=8.27$}
              \label{fig:XS_Dmax16_Pos5}
      \end{subfigure}%
      ~ 
      \begin{subfigure}[b]{0.5\columnwidth}
              \centering
              \includegraphics[trim = 0mm 0mm 0mm 0mm, clip = true,width=\textwidth]{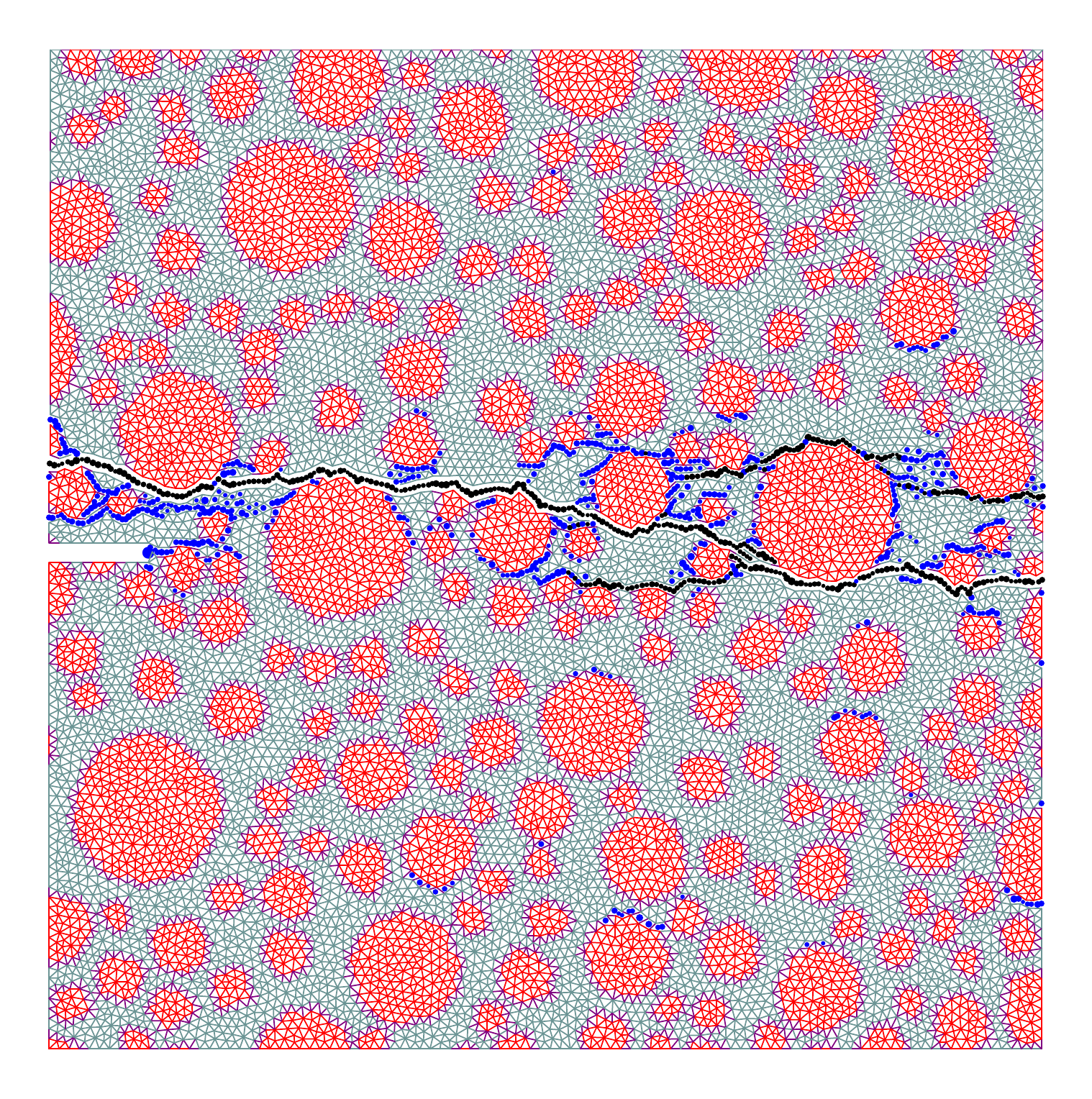}
              \caption{$d_{\max}=16$, \lFPZ{} $=16.60$}
              \label{fig:L_Dmax16_Pos5}
      \end{subfigure}%
\caption{Crack patterns and the corresponding FPZ size \lFPZ~[mm] of the XS specimens (left) and the L specimens (right), both made of the material with  $d_{\max}=6.3$~mm (top) and $d_{\max}=16$~mm (bottom). The black dots indicate broken elements with the largest opening -- read as the macrocrack -- while the blue dots stand for the remaining broken elements -- read as the microcrack.}\label{fig:L_XS_FullerDmaxP45}
\end{figure}

\Cref{fig:L_XS_FullerDmaxP45} shows the crack patterns (selected among several realizations of inclusion positions) and the value of \lFPZ{} corresponding to the smallest inclusion sizes ($d_{\max}=6.3$~mm) and the biggest ones ($d_{\max}=16$~mm) for the two extreme ligament lengths (XS and L). In the case of XS specimens (\Cref{fig:XS_Dmax6.3_Pos4,fig:XS_Dmax16_Pos5}), whatever the maximum inclusion size, a crack without bifurcation crosses the ligament by connecting ITZ elements with a path that seems to be the shortest. Whereas, in the case of L specimens (\Cref{fig:L_Dmax6.3_Pos1,fig:L_Dmax16_Pos5}), even if only one crack finally crosses the ligament, a number of microcracks occur either side of the inclusions. As a consequence, the FPZ size is in direct proportion with the maximum inclusion size in the latter case. This should be simply because the microcracks have enough space to develop in specimens with large ligament.


\subsection{Material characteristic length \textit{versus} FPZ size}
\label{sec:lc_vs_lFPZ}

The influence of the material mesostructure on the FPZ size has been studied. The aim is now to question whether the same influence can be observed on the characteristic length of the material. Although many simulations were carried out to answer this question, we only focus herein on two mesoscopic features that may influence the characteristic length \lc, the inclusion size with \begin{inparaenum}[(i)] \item fixed position and \item fixed surface fraction\end{inparaenum}.

\begin{figure}[!htbp]
 \centering
\includegraphics[width=\columnwidth]{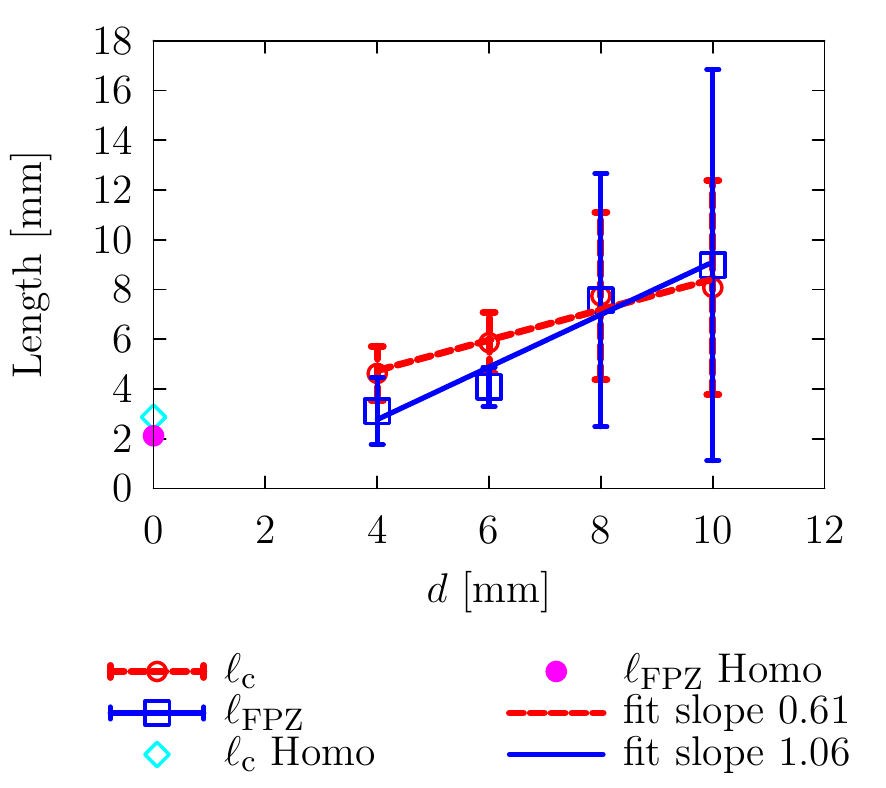}
\caption{Variation of the characteristic length of the material \lc{} and the FPZ size \lFPZ{} with respect to the inclusion size $d$ within the path (\textit{a}) of variation.}
\label{fig:lc_lFPZ_N55}
\end{figure}

First, the lattice simulations are performed by varying the inclusion size while both the positions and the number of inclusions remain unchanged. This concerns the path (\textit{a}) of mesostructure variation in which the monodisperse diameter of inclusions is changed by setting their values to $4$, $6$, $8$, and then $10$~mm. \Cref{fig:lc_lFPZ_N55} shows the relation between the characteristic length \lc{} and the inclusion size $d$. For a comparison with the FPZ size \lFPZ, the relation between \lFPZ, computed from the LD tests, and the inclusion size is shown as well. \emph{The main observation is that \lc{} and \lFPZ{} have the same order of magnitude and trend with respect to the inclusion size}. 

The increase in $d$ results in an increase of standard deviations of \lc{} as previously observed in the variation of the FPZ size \lFPZ. This is explained by the same reasons mentioned above for the FPZ size. 

Now the question is whether the variation of the characteristic length \lc{} with respect to the inclusion size $d$ still follows the variation of the FPZ size \lFPZ{} with respect to $d$ if we only do vary the size $d$ of inclusions while keeping the surface fraction of inclusions as constant as possible? For this end, the path (\textit{b}) of the mesostructure variation is used to study the influence of the inclusion size $d$ on the characteristic length \lc, in which the ``reference'' surface fraction of inclusions is kept at $45$\% when changing the inclusion size.

\begin{figure}[!htbp]
 \centering
\includegraphics[width=\columnwidth]{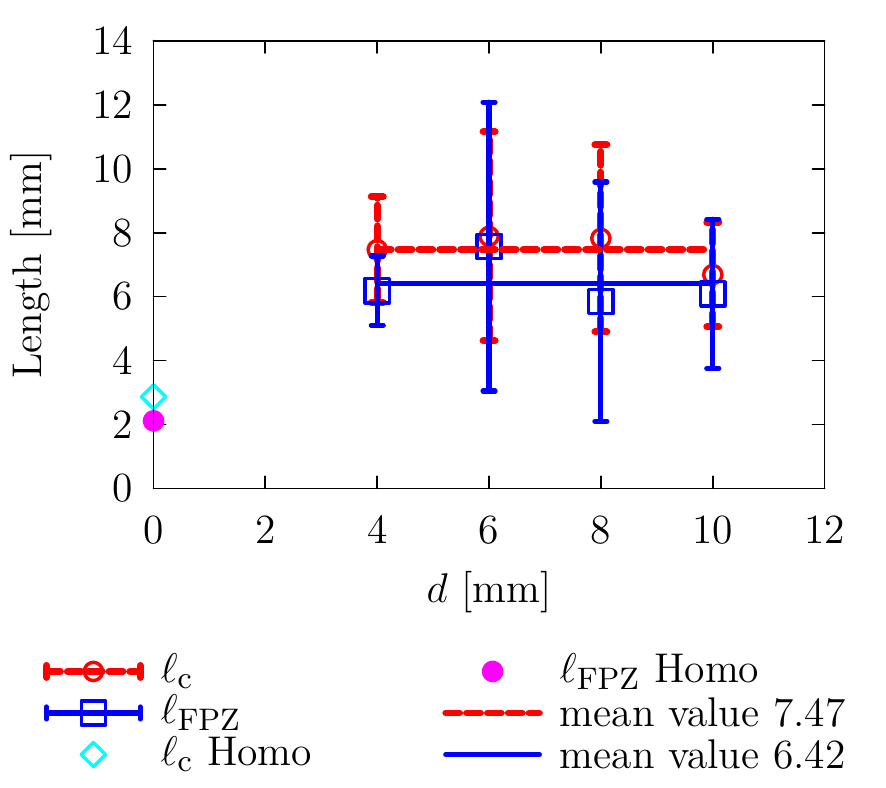}
\caption{Variation of the characteristic length of the material \lc{} and the FPZ size \lFPZ{} with respect to the inclusion size $d$ within the path (\textit{b}) of variation.}
\label{fig:lc_lFPZ_P45}
\end{figure}

\Cref{fig:lc_lFPZ_P45} shows the characteristic length of the material \lc{} as a function of the inclusion size $d$. The plot between the FPZ size \lFPZ{} and $d$ is shown as well. It exhibits that increasing $d$ does not lead to an increase of \lc, as previously observed in the case of \lFPZ. With a fixed value of the inclusion size, the resulting characteristic length of the material varies upon the spatial distribution of inclusions. However, the mean value of the characteristic length with respect to the spatial distribution of inclusions seems to be unchanged upon the increase of the inclusion size. The reason for this non-sensitivity may be related to the fact that the spacing between the inclusions, thus the spacing between the ITZs, seems to be insignificantly changed when the inclusion size is increased, as previously shown for the case of the FPZ size.

\section{Conclusion}

Two types of tensile tests were performed to study the key features that influence the FPZ size \lFPZ{} and the material characteristic length \lc{}. The assessment of \lFPZ{} is achieved \via{} localized damage (LD) tests while \lc{} is measured \via{} both LD tests and distributed damage (DD) tests. The numerical simulations are performed on the brittle elastic model material with inclusions. The material is then modeled as the three-phase material with the inclusion and matrix phases and the interfacial transition zone (ITZ) in-between them. Not only the mesostructure of the material but also the specimen geometry and the ligament size are varied in order to analyze their effect on the resulting FPZ size and material characteristic length. Five independent realizations of inclusion positions are generated for each case of the mesostructure so that the average values of \lFPZ{} and of \lc{} over that five realizations are used to analyze the effect of the mesostructure. The study points out the influences of: the inclusion size with fixed surface fraction, the inclusion size with the number and the position of inclusions unchanged, the inclusion surface fraction with fixed size, and the ligament size of the specimen on \lFPZ{} and \lc.

From that extensive study, the following conclusions can be drawn:
\begin{itemize}
 \item It appears that not basically the size, but other parameters that characterize the inclusion structure of the material such as the surface fraction, the fabric, the connectivity\dots strongly affects the size of the FPZ, and thus the characteristic length of the material.
 \item The measured value of the FPZ size is also dependent on the specimen geometry and the ligament size of specimens. Therefore, it is difficult to avoid the conclusion that the FPZ size is \emph{not} an intrinsic property of the material as usually believed. However, it seems true that the FPZ size remains in the same order when the tested system is the same (mesostructure, global geometry and dimensions, loading conditions\dots).
 \item The assessment of the characteristic length of the material is essential for using its value as the internal length in nonlocal models. However, just like the FPZ size, it is difficult to avoid structural effects in the method of measurement of the characteristic length.
\end{itemize}

This is a first step to study the influence of inclusion properties on the characteristic length and several interesting qualitative conclusions on the numerical model material have already been pointed out. For the future work, we plan to study the effect of the mechanical properties, especially the ratios of the different stiffnesses and strengths of the material phases, on the resulting FPZ size and material characteristic length. Future developments will also aim to develop the numerical model to be more representative of quasi-brittle materials, especially concrete.



\section*{Acknowledgements}

The authors gratefully acknowledge the financial support received from the BQR of Grenoble INP (contact reference: 2009/A14) for our work in 3SR laboratory. The laboratory 3SR is part of the LabEx Tec 21 (Investissements d'Avenir - grant agreement n\textsuperscript{o}ANR-11-LABX-0030).



\bibliographystyle{elsarticle-num}
\bibliography{Refs_BUI}


\end{document}